\documentclass[a4paper,12pt]{article}
\usepackage[english]{babel}
\usepackage{amssymb,amsmath,amsfonts,amsthm,epsf}
\usepackage{hyphenat}
\usepackage{latexsym}
\usepackage{graphicx}
\usepackage{feynmf}
\usepackage{graphicx,epsfig}
\usepackage{simplewick}
\usepackage{slashed}
\usepackage[retainorgcmds]{IEEEtrantools}
\usepackage{color}
\usepackage{setspace}

\numberwithin{equation}{section}

\usepackage[title,titletoc,toc]{appendix}
\usepackage{subfigure}
\usepackage{color}

\newcommand{\be}{\begin{equation}}
\newcommand{\ee}{\end{equation}}
\newcommand{\bd}{\begin{definition}}
\newcommand{\ed}{\end{definition}}
\newcommand{\bt}{\begin{theorem}}
\newcommand{\et}{\end{theorem}}
\newcommand{\bp}{\begin{proof}}
\newcommand{\ep}{\end{proof}}
\newcommand{\bea}{\begin{eqnarray}}
\newcommand{\eea}{\end{eqnarray}}
\newcommand{\ba}{\begin{array}}
\newcommand{\ea}{\end{array}}

\def\siml{{\ \lower-1.2pt\vbox{\hbox{\rlap{$<$}\lower6pt\vbox{\hbox{$\sim$}}}}\ }} 
\def\bfnabla{\mbox{\boldmath $\nabla$}}

\def\bfsigma{\mbox{\boldmath $\sigma$}}
\def\bftau{\mbox{\boldmath $\tau$}}

\def\al{\alpha}
\def\lQ{\Lambda_{\rm QCD}}

\newcommand{\nn}{\nonumber}

\def\bfnabla{\mbox{\boldmath $\nabla$}}

\newcommand{\MS}{\overline{\rm MS}}

\newcommand{\eq}[1]{Eq.~(\ref{#1})}
\newcommand{\fig}[1]{Fig.~\ref{#1}}

\def\dsl{\,\raise.15ex\hbox{/}\mkern-13.5mu D}

\def\I{\rm 1\kern-.24em l}  % Yes, I know. This ain't capital.

\newcommand{\MeV}{\,\mathrm{MeV}}

\newcommand{\mpi}{m_\pi}
\newcommand{\vq}{{\bf  q}}
\newcommand{\mt}{\tilde m}
\newcommand{\tL}{\tilde L}
\newcommand{\intx}{\int_0^1 dx\,}
\newcommand{\qv}{q_0}

\newcommand{\beq}{\begin{equation}}
\newcommand{\eeq}{\end{equation}}

\newcommand{\Appendix}[1]%
    {%
     \section{#1}%
      }

\numberwithin{equation}{section}

\begin{document}

\begin{titlepage}
\begin{flushright}
\end{flushright}

\vspace{1cm}
\begin{center}
\begin{Large}
{\bf 
The two-photon exchange contribution to muonic hydrogen from chiral perturbation theory
%Matching HBET to NRQED: the four fermion sector
%and the forward virtual Compton tensor
}\\[2cm] 
\end{Large} 
{\large Clara Peset and Antonio Pineda}\\
\vspace{0.5cm}
{\it  Grup de F\'\i sica Te\`orica, Universitat Aut\`onoma de Barcelona,\\ 
E-08193 Bellaterra (Barcelona), Spain}\\
\today
\end{center}

\vspace{1cm}

\begin{abstract}
We compute the spin-dependent and spin-independent structure functions of the forward virtual-photon Compton tensor 
of the proton at ${\cal O}(p^3)$ using heavy baryon effective theory including the Delta particle. We compare with previous results when existing.
Using these results we obtain the leading hadronic contributions, associated to the pion and Delta 
particles, to the Wilson coefficients of the lepton-proton four fermion operators in NRQED. 
The spin-independent coefficient yields a pure prediction for the two-photon 
exchange contribution to the muonic hydrogen Lamb shift, $\Delta E_{\rm TPE}(\pi\&\Delta)=34(13)$ $\mu$eV. We also compute the charge, $\langle r^{n} \rangle$, 
and Zemach, $\langle r^{n} \rangle_{(2)}$, moments for $n \geq 3$. Finally, we discuss 
the spin-dependent case, for which we compute the difference between the four-fermion Wilson coefficients
 relevant for hydrogen and muonic hydrogen.
\vspace{5mm} \\
PACS numbers:12.39.Fe, 11.10.St, 12.20.Ds, 12.39.Hg
\end{abstract}

\end{titlepage}
\vfill
\setcounter{footnote}{0} 
\vspace{1cm}

\section{Introduction}
The spin-dependent and spin-independent structure functions of $T^{\mu\nu}$, the forward virtual-photon Compton tensor 
of the proton, carry important information about the QCD dynamics. They test the Euclidean region of the 
theory since $Q^2\equiv -q^2 > 0$. For 
$Q^2 \sim m_{\pi}^2 \not=0$, the behavior of $T^{\mu\nu}$ is determined by the chiral theory, and can be obtained 
within a chiral expansion using Heavy Baryon Effective Theory (HBET) \cite{Jenkins:1990jv}. If one works 
within a large $N_c$ ideology (where $N_c$ is the number of colours) the Delta particle should be incorporated in the HBET Lagrangian~\cite{Dashen:1993ac}, 
as the Delta and the nucleon become degenerate in the large $N_c$ limit. We use this motivation to incorporate the Delta particle in the effective Lagrangian. We do so along the lines of Refs.~\cite{Jenkins:1991es,Hemmert:1996xg,Hemmert:1997ye}, i.e. we do not impose the large $N_c$ relations among the couplings but let them free and fit to the data. 
This effective field theory has a double expansion in $ \sim m_{\pi}/m_{\rho}$ 
and $\sim \Delta/m_{\rho}$, where $\Delta=M_{\Delta}-M_N$. Note that this creates a new expansion parameter $m_{\pi}/\Delta \sim 1/2$;
the associated corrections will be incorporated in our computation together with the pure chiral result. 

Within this framework we compute the spin-dependent and spin-independent structure functions of the forward virtual-photon Compton tensor 
of the proton at ${\cal O}(p^3)$ in Heavy Baryon Chiral Perturbation Theory (HB$\chi$PT) including the Delta particle. $T^{\mu\nu}$ cannot be directly related to cross sections obtained at fixed energies, as it tests the Euclidean regime. 
Nevertheless, it is possible to obtain it (up to eventual subtractions) from experiment through dispersion relations, 
i.e., through specifically weighted averages of measured cross sections over all energies. Possible constructions 
are the so-called generalized sum rules, which, for large energies, can be related with the deep inelastic sum rules. 
These have been studied in Ref.~\cite{Ji:1999mr} for the spin-dependent case. The spin-independent case has been briefly 
discussed in Ref.~\cite{Nevado:2007dd}. We will not enter into this interesting line of research in this paper. 

Instead, our main motivation for obtaining the chiral structure of $T^{\mu\nu}$ is that $T^{\mu\nu}$ 
appears in the matching computation between HBET and non-relativistic QED (NRQED) that determines 
 $c^{pl_i}_3$ and $c^{pl_i}_4$ ($l_i=e$ or $\mu$), the Wilson coefficients of the 
lepton-proton four-fermion operators in the NRQED \cite{Caswell:1985ui} Lagrangian. As soon as hadronic effects start to become important in atomic physics, these Wilson coefficients play a major role.
They appear in the hyperfine splitting (spin-dependent) and Lamb shift (spin-independent) in hydrogen and muonic hydrogen 
(see Refs.~\cite{Pineda:2002as,Pineda:2004mx,Nevado:2007dd}).  
Therefore, their determination allows us to relate the energy shifts obtained in hydrogen and muonic hydrogen. 
Even more important, these Wilson coefficients usually carry most of the theoretical uncertainty in these splittings.
This is particularly so in the case of the muonic hydrogen Lamb shift. At present, it is the limiting factor for improving the precision of the 
determination of the electromagnetic proton radius from the measurements taking place at PSI~\cite{Pohl:2010zza,Antognini:1900ns} of the muonic hydrogen spectra. 
This necessity to improve our knowledge (of the spin-independent) lepton-proton four-fermion Wilson coefficient has led 
us to compute this quantity in HB$\chi$PT including the Delta particle. Fortunately enough, this object is chiral enhanced. Therefore, the ${\cal O}(p^3)$ chiral computation yields a pure prediction, without the need of new counterterms, of $\Delta E_{\rm TPE}$, the (hadronic) two-photon exchange contribution to the muonic hydrogen
Lamb shift: $\Delta E_{\rm L}=E(2P_{3/2})-E(2S_{1/2})$. Note that, since $m_{\mu}/m_{\pi} \sim 1$, we keep the complete  $m_{\mu}/m_{\pi}$ dependence in such predictions. 
These results have been used in the recent determination of the muonic hydrogen Lamb shift and the proton radius performed in Ref.~\cite{Peset:2014yha}. One of the main motivations of this paper is to give the details of the hadronic-related part of that analysis.

We profit this analysis to revisit the distinction between the Born and non-Born terms of $T^{\mu\nu}$ and $\Delta E_{\rm TPE}$. Such distinction produces the so-called Zemach (or Born) and polarizability corrections to 
the Wilson coefficients (names also used for the associated contributions to the energy shifts: hyperfine or Lamb shift). For the spin-independent case we have a good 
analytical control and can also compute the charge, $\langle r^{n} \rangle$, and 
the Zemach, $\langle r^{n} \rangle_{(2)}$, moments, for $n\geq 3$, since they are dominated by the chiral theory. The polarizability correction of $\Delta E_{\rm TPE}$ is also usually split into the so-called inelastic and subtraction terms. We will also discuss what HB$\chi$PT has to say in this respect. 

The paper is distributed in the following way. In Sec. \ref{Sec:EFT} we present HBET and NRQED. In  Sec. \ref{Sec:tensor}  we compute $T^{\mu\nu}$. 
In Sec.~\ref{Sec:c3} we compute $c^{pl_i}_3$, $\langle r^{2k+1} \rangle$, 
and $\Delta E_{\rm TPE}$. For the latter we also discuss its separation into Born, polarizability, inelastic and subtraction terms. In Sec.~\ref{Sec:c4} we discuss about $c_4^{pl_i}$ 
and the Zemach radius, $\langle r_Z \rangle$, before we conclude. 

\section{Effective Field Theories}
\label{Sec:EFT} 

In this section, we will present the main building blocks of the HBET and NRQED Lagrangians needed for our analysis
(see also Ref.~\cite{Pineda:2002as}).

\subsection{HBET}
\label{secHBET}

Our starting point is the SU(2) version of HBET coupled to leptons 
where the Delta particle is kept as an explicit degree of freedom.  The degrees
of freedom of this theory are the proton, neutron and Delta, for which
the NR approximation can be taken, and pions, leptons
(muons and electrons) and photons, which will be taken relativistic.
This theory has a cut-off $\mu << M_p$,
$m_{\rho}$, which is much larger than any other scale in the problem.
The Lagrangian can be split in several sectors. Nevertheless, the fact that some particles will only enter through
loops, since only some specific final states are wanted, simplifies
the problem. The Lagrangian can be written as an expansion in $e$ and $1/M_p$
and can be structured as follows
\be
\label{LHBET}
{\cal L}_{HBET}=
{\cal L}_{\gamma}
+
{\cal L}_{l}
+
{\cal L}_{\pi}
+
{\cal L}_{l\pi}
+
{\cal L}_{(N,\Delta)}
+
{\cal L}_{(N,\Delta)l}
+
{\cal L}_{(N,\Delta)\pi}
+
{\cal L}_{(N,\Delta)l\pi},
\ee
representing the different sectors of the theory. In particular, the
$\Delta$ stands for the Delta particle: the spin 3/2 baryon multiplet (we also use
$\Delta=M_{\Delta}-M_p$, the specific meaning in each case should be
clear from the context).

The photonic Lagrangian reads (the first corrections to this expression 
scale like $\al^2/M_p^4$)
\be
\label{Lg}
{\cal L}_\gamma=-\frac{1}{4}F^{\mu\nu}F_{\mu \nu} 
+\left({d_{2,R} \over M_p^2}+{d_{2}^{(\tau)} \over m_{\tau}^2} 
\right) F_{\mu \nu} D^2 F^{\mu \nu}
\,,
\ee
where $d_{2,R}$ stands for the hadronic contribution. 
The second term will not be considered any further in this paper, 
since we are mainly interested in the lepton-proton four-fermion operators.

The leptonic sector can be approximated to ($i D_\mu=i\partial_\mu-eA_\mu$)
\be
\label{Ll}
{\cal L}_l=\sum_i \bar l_i  (i\dsl-m_{l_i}) l_i
\,,
\ee
where $l_i=e,\mu$. 

The Lagrangian of a heavy baryon at ${\cal O}(1/M_p^2)$ coupled to electromagnetism reads
\bea
\label{LNdelta}
 {\cal L}_{N}&=& N^\dagger_{p} \Biggl\{iD_0+ {{\bf
D}^2_p\over 2 M_p} + 
 {{\bf D}_p^4\over
8 m^3_{p}} - e {c_F^{(p)} \over 2M_p}\, {\bf \bfsigma \cdot B}
\\
&& 
\nn
-e{c_D^{(p)} \over 8M_p^2}
   \left[{\bf \bfnabla \cdot E }\right] 
 - ie { c_S^{(p)} \over 8M_p^2}\, \bfsigma \cdot \left({\bf D}_p
     \times {\bf E} -{\bf E}    \times {\bf D}_p\right) 
\Biggr\} N_{p}
\,,
\eea
where $ i D^0_p=i\partial_0 +Z_peA^0$, $i{\bf D}_p=i{\bfnabla}-Z_pe{\bf A}$. 
For the proton $Z_p=1$ (for the neutron $Z_p=0$ and for all indices $p \rightarrow n$).
 
The Delta particle mixes with the nucleons at
${\cal O}(1/M_p)$ (${\cal O}(1/M_p^2)$ terms are not needed in our case). The only
relevant interaction in our case is the $p$-$\Delta^+$-$\gamma$ term,
which is encoded in the second term of 
\be {\cal L}_{(N,\Delta)}
= 
T^{\dagger}(i\partial_0-\Delta)T
+{eb_{1,F} \over 2M_p}
\left(
T^{\dagger}\bfsigma ^{(3/2)}_{(1/2)}\cdot {\bf
B}\,\bftau^{3(3/2)}_{(1/2)} N + h.c.
\right)
\,,
\ee  
where $T$ stands for the delta 3/2 isospin multiplet, $N$ for the nucleon 1/2
isospin multiplet and the transition spin/isospin matrix elements fulfill
(see \cite{Weise})
\be
\bfsigma^{i(1/2)}_{(3/2)}\bfsigma^{j(3/2)}_{(1/2)}
={1 \over 3}(2\delta^{ij}-i\epsilon^{ijk}\bfsigma^k),
\qquad
\bftau^{a(1/2)}_{(3/2)}\bftau^{b(3/2)}_{(1/2)}
={1 \over 3}(2\delta^{ab}-i\epsilon^{abc}\bftau^c).
\ee

The baryon-lepton Lagrangian provides new terms that are not usually considered 
in HBET. The relevant term in our case is the interaction between 
the leptons and the nucleons (actually only the proton):
\be
\label{LNl}
 {\cal L}_{(N,\Delta)l}=\displaystyle\frac{1}{M_p^2}\sum_i c_{3,\rm R}^{pl_i}
{\bar N}_p \gamma^0
  N_p \ \bar{l}_i\gamma_0 l_i
+\displaystyle\frac{1}{M_p^2}\sum_i c_{4,\rm R}^{pl_i}{\bar N}_p \gamma^j \gamma_5
N_p 
  \ \bar{l}_i\gamma_j \gamma_5 l_i
\,.
\ee
The above matching coefficients fulfill 
$c_{3,\rm R}^{pl_i}=c_{3,\rm R}^{p}$ and $c_{4,\rm R}^{pl_i}=c_{4,\rm R}^{p}$ up to terms
suppressed by $m_{l_i}/M_p$, which will be sufficient for our purposes. 

Let us note that with the conventions above, $N_p$ is the field of the proton 
(understood as a particle) with
positive charge if $l_i$ represents the leptons (understood as
particles) with negative charge.

The hadronic interactions are organized according to their chiral counting. 
Since a single chiral loop already produces a factor $1/(4\pi F_0)^2 \sim
1/M_p^2$, we only need the leading pionic Lagrangian coupled to electromagnetism:
\be
{\cal L}_\pi=\left[(\partial_\mu-ieA^\mu)\pi^+\right]\left[(\partial^\mu+ieA^\mu) \pi^-\right]-m_\pi^2\pi^+\pi^-
+{1 \over 2}(\partial_\mu \pi^0)(\partial^\mu \pi^0)-{1 \over
2}m_\pi^2\pi^0\pi^0
\,.
\ee
We do not need to account for pion self-interactions, and
the pion-baryon interactions are only needed at
${\cal O}(m_\pi)$, the leading order, which is known
\cite{Jenkins:1991es,Bernard:1992qa,Bernard:1995dp,Hemmert:1996rw}:
\be
{\cal L}_{ (N,\Delta)\pi}= \bar N \left(i \Gamma_0+g_A u\cdot S\right)N+g_{\pi N\Delta}\left(\bar T_a^\mu w_\mu^a N+h.c.\right)
\ee
where
\bea
U&=&u^2=e^{i {\bf {\tau\cdot \pi}}/F_{\pi}},\\
D_\mu&=&\partial_\mu+\Gamma_\mu,\\
\Gamma_\mu&=&\frac{1}{2}\left\{u^\dagger\partial_\mu u+u\partial_\mu u^\dagger-i\frac{e}{2}A_\mu\left(u^\dagger\tau^3 u+u\tau^3 u^\dagger\right)\right\},\\
u_\mu&=&iu^\dagger\nabla_\mu U u^\dagger,\\
w_\mu^a&=&\frac{1}{2}Tr[\tau^a u_\mu]=-\frac{1}{F_\pi}\partial_\mu \pi^a-\frac{e}{F_\pi}A_\mu \epsilon^{a3b}\pi^b+...
\,.
\eea
$T_a^\mu $ is the Rarita-Schwinger spin 3/2 field and 
$S_{\mu}=\frac{i}{2} \gamma_5 \sigma_{\mu\nu}v^\nu$ is the spin operator 
(where we take $v_{\mu}=(1,{\bf 0})$).

This finishes all the needed terms for this paper, since the other sectors
of the Lagrangian would give subleading contributions.
 
\subsection{NRQED($\mu$)}
\label{NRQED(mu)}

In the muon-proton sector, by integrating out the $m_\pi$ and $\Delta$ scales, an
effective field theory for muons, protons and photons appears. In
principle, we should also consider neutrons but they play no role at
the precision we aim. The effective theory corresponds to a hard
cut-off $\nu << m_\pi$ and therefore pions and Deltas have been
integrated out. The Lagrangian is equal to the previous case but
with neither pions nor Deltas, and with the following modifications: ${\cal
L}_{l} \rightarrow {\cal L}_{e}+{\cal L}^{(\rm NR)}_\mu$ and ${\cal
L}_{(N,\Delta)l} \rightarrow {\cal L}_{Ne}+{\cal L}^{(\rm NR)}_{N\mu}$,
where it is made explicit that the the muon has become NR. Any further difference goes into the Wilson 
coefficients, in particular, into the Wilson coefficients of the
baryon-lepton operators. In summary, the Lagrangian reads
\be
{\cal L}_{\rm NRQED(\mu)}=
{\cal L}_{\gamma}
+
{\cal L}_{e}
+
{\cal L}^{(\rm NR)}_{\mu}
+
{\cal L}_{N}
+
{\cal L}_{Ne}
+
{\cal L}_{N\mu}^{(\rm NR)}
\,,
\ee
where
\bea
\label{LNdeltaNR}
 {\cal L}_{N}&=& N^\dagger_{p} \Biggl\{iD_0+ {{\bf
D}^2_p\over 2 M_p} + 
 {{\bf D}_p^4\over
8 m^3_{p}} - e {c_F^{(p)} \over 2M_p}\, {\bf \bfsigma \cdot B}
\\
&& 
\nn
-e{c_D^{(p)} \over 8M_p^2}
   \left[{\bf \bfnabla \cdot E }\right] 
 - ie { c_S^{(p)} \over 8M_p^2}\, \bfsigma \cdot \left({\bf D}_p
     \times {\bf E} -{\bf E}    \times {\bf D}_p\right) 
\Biggr\} N_{p}
\,,
\eea 
\be
{\cal L}^{(\rm NR)}_{\mu}= l_{\mu}^\dagger \Biggl\{ i D_{\mu}^0 + \, {{\bf D}_{\mu}^2\over 2 m_{\mu}} +
 {{\bf D}_{\mu}^4\over
8 m^3_{\mu}} + eZ_{\mu} {c_F^{(\mu)} \over 2m_\mu}\, {\bf \bfsigma \cdot B}
 +
 i eZ_{\mu} {c_S^{(\mu)} \over 8 m^2_{\mu}}\,  {\bf \bfsigma \cdot \left(D_{\mu} \times
     E -E \times D_{\mu}\right) } 
\Biggr\} l_{\mu}
\ee
and\footnote{$c_{3/4}^{pl_\mu} \rightarrow c_{3/4,\rm NR}^{pl_\mu}$ in Ref.~\cite{Pineda:2002as}. 
We eliminate some subindeces to lighten the notation.}
\be
{\cal L}_{N\mu}^{\rm NR}=
\displaystyle\frac{c_{3}^{pl_{\mu}}}{M_p^2} N_p^{\dagger}
  N_p \ {l}^{\dagger}_\mu l_\mu
-\displaystyle\frac{c_{4}^{pl_{\mu}}}{M_p^2} N_p^{\dagger}{\bfsigma}
  N_p \ {l}^{\dagger}_\mu{\bfsigma} l_\mu
\,,
\ee
with the following definitions: $ i D^0_\mu=i\partial_0 -Z_\mu eA^0$,
$i{\bf D}_\mu=i{\bfnabla}+Z_\mu e{\bf A}$ and $Z_\mu=1$. ${\cal L}_e$
stands for the relativistic leptonic Lagrangian in Eq. (\ref{Ll}) and ${\cal
L}_{Ne}$ for Eq. (\ref{LNl}), both for the electron case only.

\begin{figure}[h]
%\makebox[2.0cm]{\phantom b}
%\epsfxsize=9truecm \epsfbox{polhf.eps}
\begin{center}
\includegraphics[width=0.35\textwidth]{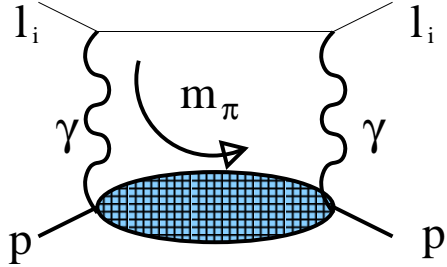}
\caption {\it Symbolic representation of the matching between HBET and NRQED for $c_3^{pl_i}$ and $c_4^{pl_i}$. The bubble 
represents the hadronic corrections.}
\label{fig:c3}
\end{center}
\end{figure}

Our main interest is the determination of $c_{3}^{pl_\mu}$ and $c_{4}^{pl_\mu}$ by matching 
HBET to NRQED. At ${\cal O}(\al^2)$ we can symbolically represent this matching as in Fig.~\ref{fig:c3}.

\subsection{NRQED($e$)}

If we focus in the electron-proton sector, things go quite as in the previous section. 
After integrating out scales of ${\cal O}(m_\pi,\Delta)$, an effective field theory for electrons coupled to protons
(and photons) appears.  This effective theory has a
cut-off $\nu << m_\pi$ and pions, Deltas and muons have been
integrated out, but the electron is still relativistic. 
After integrating out scales of ${\cal O}(m_e)$ in the electron-proton
sector, we still have an effective field theory for electrons coupled
to protons and photons.  Nevertheless, now the electrons are
NR. The Lagrangian is quite similar to the one in
Subsec. \ref{NRQED(mu)} but without a light fermion and with the
replacement $\mu \rightarrow e$. It reads
\be
{\cal L}_{NRQED(e)}=
{\cal L}_{\gamma}
+
{\cal L}^{(NR)}_{e}
+
{\cal L}_{N}
+
{\cal L}_{Ne}^{(NR)}
\,.
\ee
We will perform the matching to this theory directly from HBET.  
At ${\cal O}(\al^2)$ this matching can be symbolically represented by the same figure as in the case of the muon, 
namely Fig.~\ref{fig:c3}.

\section{Forward virtual Compton tensor $T^{\mu\nu}$}
\label{Sec:tensor}

The electromagnetic current reads $J^\mu=\sum_i Q_i{\bar q}_i\gamma^\mu q_i$, where $i=u,d$ (we will 
not consider the strange quark in this paper)
and $Q_i$ is the quark charge. The 
form factors (which we will understand 
as pure hadronic quantities, i.e. without electromagnetic 
corrections) are then defined by the following equation:   
\be
\langle {p^\prime,s}|J^\mu|{p,s}\rangle
=
\bar u(p^\prime) \left[ F_1(q^2) \gamma^\mu +
i F_2(q^2){\sigma^{\mu \nu} q_\nu\over 2 M_p} \right]u(p)
\label{current}
\,,
\ee
where $q=p'-p$  and $F_1$, $F_2$ are the Dirac and Pauli form factors,
respectively. The states
are normalized in the following (standard relativistic) way:
\be
\langle p',\lambda'|p,\lambda \rangle
=
(2\pi)^3 2p^0\delta^3({\bf p}'-{\bf p})
\delta_{\lambda' \lambda}\,,
\ee
and 
\be
u(p,s){\bar u}(p,s)=({\rlap/p+M_p }){1+\gamma_5{\rlap/s} \over 2}
\,,
\ee
where $s$ is an arbitrary spin four-vector obeying $s^2=-1$ and
$p\cdot s=0$.
 
\begin{figure}[h]
\begin{center}
\includegraphics[width=0.65\textwidth]{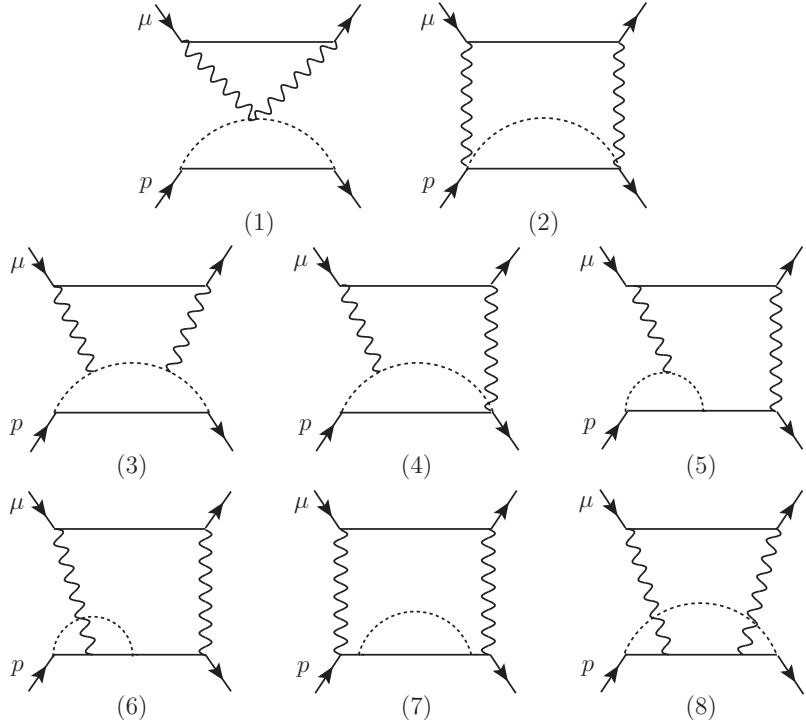}
\caption {\it Two-loop diagrams with an internal pion loop contributing to $c^{pl_{\mu}}_3$ and $c_4^{pl_{\mu}}$. Crossed diagrams and those obtained through permutations are implicit.}\label{piL}
\end{center}
\end{figure}

\begin{figure}[h]
\begin{center}
\includegraphics[width=0.3\textwidth]{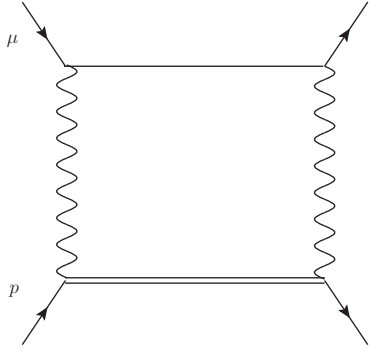}
\caption{\it One-loop diagram with an internal Delta particle contributing to $c_3^{pl_{\mu}}$ and $c_4^{pl_{\mu}}$. Crossed diagram is implicit.
}\label{DTL}
\end{center}
\end{figure}

\begin{figure}[h]
\begin{center}
\includegraphics[width=0.65\textwidth]{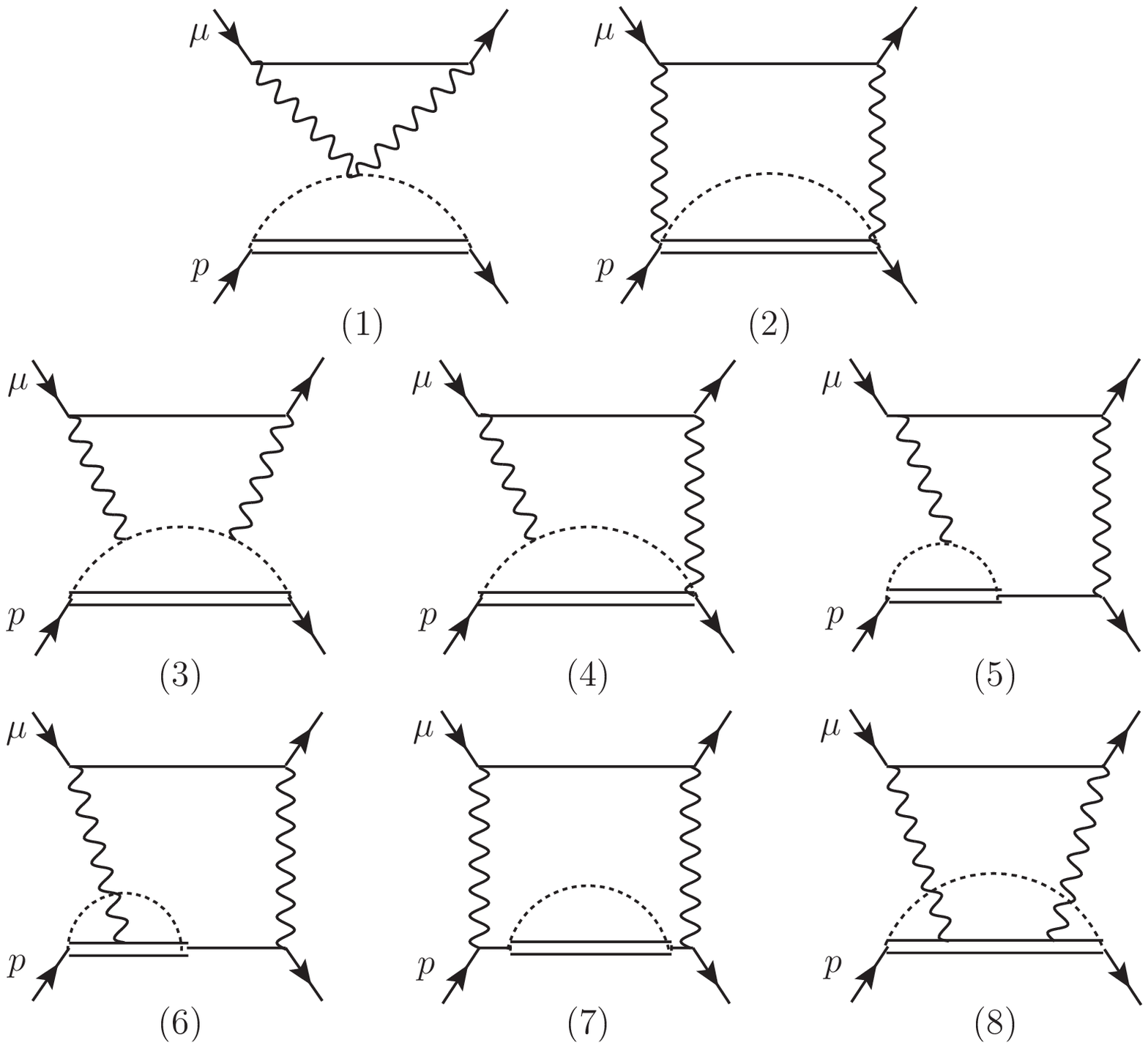}
\caption {\it  Two-loop diagrams with an internal pion and Delta loop contributing to $c_3^{pl_{\mu}}$ and $c_4^{pl_{\mu}}$. Crossed diagrams and those obtained through permutations are implicit.}\label{D1L}
\end{center}
\end{figure}

More suitable for a NR analysis are the Sachs form factors:
\be
G_E(q^2)=F_1(q^2)+{q^2 \over 4M_p^2}F_2(q^2), \qquad G_M(q^2)=F_1(q^2)+F_2(q^2). 
\ee
Nevertheless, the main object of interest of this paper is the forward virtual-photon Compton tensor,
\begin{equation} 
 T^{\mu\nu} = i\!\int\! d^4x\, e^{iq\cdot x}
  \langle {p,s}| T \{J^\mu(x)J^\nu(0)\} |{p,s}\rangle
\,,
\end{equation}
which has the following structure ($\rho=q\cdot p/M_p\equiv v \cdot q$, although we will usually work in the 
rest frame where $\rho=q^0$):
\bea \label{inv-dec}
 T^{\mu\nu} &=
  &\left( -g^{\mu\nu} + \frac{q^\mu q^\nu}{q^2}\right) S_1(\rho,q^2) 
  %\nn\\*& 
  + \frac1{M_p^2} \left( p^\mu - \frac{M_p\rho}{q^2} q^\mu \right)
    \left( p^\nu - \frac{M_p\rho}{q^2} q^\nu \right) S_2(\rho,q^2) 
	\\*
  && - \frac i{M_p}\, \epsilon^{\mu\nu\rho\sigma} q_\rho s_\sigma A_1(\rho,q^2)
    %\nn\\*& 
	- \frac i{M_p^3}\, \epsilon^{\mu\nu\rho\sigma} q_\rho
   \bigl( (M_p\rho) s_\sigma - (q\cdot s) p_\sigma \bigr) A_2(\rho,q^2)
   \equiv T_S^{\mu\nu}+T_A^{\mu\nu}
   \nn
   \,.
\eea
It depends on four scalar functions, which we call structure functions. 
We split the tensor into the symmetric (spin-independent),
$T_S^{\mu\nu}=T_S^{\nu\mu}$ (the first two terms of Eq.~(\ref{inv-dec})), 
and antisymmetric (spin-dependent) pieces, $T_A^{\mu\nu}=-T_A^{\nu\mu}$ (the last two terms of Eq.~(\ref{inv-dec})). 
We have computed this tensor at ${\cal O}(p^3)$ in HB$\chi$PT. The diagrams that contribute are listed in Figs. \ref{piL}, 
\ref{DTL} and \ref{D1L} (without closing the loop with the muon, i.e. without the muon line).
The first figure refers to diagrams without Delta contributions (pure chiral), the second to the tree-level Delta contribution, and the 
last to one-loop chiral diagrams involving the Delta particle. Expressions in $D=4-\epsilon$ and four dimensions for each diagram 
can be found in Appendix \ref{Amp}. Summing them up we can reconstruct the tensor structure of $T^{\mu\nu}$
(in other words, check gauge invariance). 
In principle, more diagrams, besides those drawn should be considered but they do not contribute to the 
structure functions at the order we aim in this work.

It is also common to split $T^{\mu\nu}$ into two components, which we label "Born" and "pol":
\be
T^{\mu\nu}=T^{\mu\nu}_{\rm Born}+T^{\mu\nu}_{\rm pol}\,. 
\ee
The Born term is defined as the contribution coming from the intermediate state being the proton (somewhat the 
elastic contribution). The associated structure functions can be written in terms of the form factors. 
They read (or, rather, they are defined as)
\bea
 S_1^{\rm Born}(\rho,q^2) & \equiv
  -2F_1^2(q^2) - \frac {2(q^2)^2\,G_{\rm M}^2(q^2)}{(2M_p\rho)^2-(q^2)^2}, \\
 S_2^{\rm Born}(\rho,q^2) & \equiv
  2\, \frac {4M_p^2q^2\,F_1^2(q^2)-(q^2)^2\,F_2^2(q^2)}{(2M_p\rho)^2-(q^2)^2}, \\
 A_1^{\rm Born}(\rho,q^2) & \equiv  \label{Born.A1}
  -F_2^2(q^2) + \frac {4M_p^2q^2\,F_1(q^2)G_{\rm M}(q^2)}{(2M_p\rho)^2-(q^2)^2}, \\
 A_2^{\rm Born}(\rho,q^2) & \equiv
  \frac {4M_p^3\rho\,F_2(q^2)G_{\rm M}(q^2)}{(2M_p\rho)^2-(q^2)^2}.
\eea
From these expressions one could easily single out the point-like contributions. The remaining contributions,  
with the ${\cal O}(p^3)$ accuracy of our chiral computation, are encoded in the 
following expression (we split $G_{E,M}$ into pieces according to its chiral counting:
$G_{E,M}^{(n)} \sim 1/M_p^n\sim 1/\Lambda_{\chi}^n$):
\bea
\label{TBorn}
&&T^{\mu\nu}_{\rm Born}
=
i\pi\delta(v\cdot q)
\\
&&
\nn
\times
Tr\left[u\bar u
\left(
-4p_+G_E^{(0)}G_E^{(2)}v^{\mu}v^{\nu}
+\frac{2}{M_p}G_E^{(0)}G_M^{(1)}
\left(
v^{\mu}p_+\left[s^{\nu},s^{\rho'}\right]q_{\rho'}p_+
-
v^{\nu}p_+\left[s^{\mu},s^{\rho'}\right]q_{\rho'}p_+
\right)
\right)
\right]
\,,
\eea
where $p_+=\frac{1+v\cdot \gamma}{2}$. 
Note that $T^{\mu\nu}_{Born}$ is proportional to $\delta(v\cdot q)$ and 
$G_E^{(0)}=1$. The expressions for $G_E^{(2)}$, $G_M^{(1)}$ can be found in 
Refs. \cite{Bernard:1992qa,BKKM,BFHM}. We write them here for ease of reference:
\bea
G^{(2)}_E( q^2)&=&q^2\frac{\langle r^2\rangle}{6} +\frac{1}{(4\pi  F_\pi)^2}\left(q^2\left(\frac{1}{12}+\frac{g_A^2}{4}-\frac{2g_{\pi N \Delta}^2}{9}\right)\right.\nn\\
&-&\left.\frac{4 }{3 } g_{\pi N \Delta}^2 \Delta \left(\frac{5}{9}\frac{q^2}{\sqrt{\Delta^2-\mpi^2}}+4\sqrt{\Delta^2-\mpi^2}\right)\ln {\cal R}(\mpi^2)\right)\nn\\
&+&\frac{1}{(4\pi  F_\pi)^2}\intx\left\{\left[\mpi^2\left(\frac{1}{2}+\frac{3}{2} g_A^2-\frac{4}{3} g_{\pi N \Delta}^2\right)+\Delta ^2\frac{8}{3}g_{\pi N \Delta}^2\right.\right.\nn\\
&+&\left.\left.\left(\frac{1}{2}+\frac{5}{2} g_A^2-\frac{20}{9} g_{\pi N \Delta}^2\right) q^2 (-1+x) x\right] \ln\left(\frac{\mt^2}{\mpi^2}\right)\right.\nn\\
&+&\left.\frac{16}{3} g_{\pi N \Delta}^2 \frac{\Delta}{\sqrt{\Delta ^2-\mt^2}} \left(\frac{4}{3}q^2x (1-x)+ \Delta^2-\mpi^2 \right)\ln {\cal R}(\mt^2)\right\}\label{GE2}
\,,
\eea
where (the coefficients $\tilde{B_1}$ and $B_{10}$ are counterterms of the HBET Lagrangian from \cite{BFHM})
\bea
\langle r^2\rangle &=&-6\frac{dG_E(-{\bf q^2})}{d({\bf q^2})}\Bigg|_{{\bf q}^2=0} =\frac{3(\kappa_s+\kappa_v)}{4M_p^2}-\frac{1}{(4\pi  F_\pi)^2}\left(\frac{1}{2}+12\tilde{B_1}+6 B_{10}+\frac{7}{2}g_A^2-\frac{104}{27} g_{\pi N \Delta}^2 \right.\nn\\
&-&\left.\frac{40}{9}g_{\pi N \Delta}^2  \frac{\Delta }{ \sqrt{\Delta ^2-\mpi^2}}\ln\left({\cal R}(\mpi^2)\right)+\left(1+5 g_A^2-\frac{40}{9} g_{\pi N \Delta}^2\right) \ln\left(\frac{\mpi}{\lambda }\right)\right),
\eea
and
\bea
\label{GM1}
G_M^{(1)}(q^2)&=&- g_{A}^2 \frac{4\pi M_p}{(4\pi F_\pi)^2} 
                   \int_{0}^{1}dx\left\{ \sqrt{\tilde{m}^2}-m_{\pi}\right\}
                    \\
                & &+\frac{32}{9}g_{\pi N\Delta}^2\frac{
                   M_p \Delta}{
                   (4\pi F_\pi)^2}\int_{0}^{1}dx \left\{ \frac{1}{2}\ln\left(
                   \frac{\tilde{m}^2}{4\Delta^2}\right)-\ln\left(\frac{m_\pi}{
                   2\Delta}\right) \right. \nonumber \\
                & &\phantom{+\frac{32}{9}g_{\pi N\Delta}^2\frac{
                   M_N \Delta}{(4\pi F_\pi)^2}\int_{0}^{1}dx} \left. 
                   +\frac{\sqrt{\Delta^2-\tilde{m}^2}}{\Delta}
                   \ln {\cal R}\left(\tilde{m}^2\right)
                   -\frac{\sqrt{\Delta^2-m_{\pi}^2}}{
                   \Delta}\ln {\cal R}(\mpi^2)\right\} \; ,
\nn
\eea

with 
\begin{equation}
{\cal R}\left(m^2\right)
= \frac{\Delta}{m}+\sqrt{\frac{\Delta^2}{m^2}-1}~, \quad
\tilde{m}^2 = m_\pi^2 - q^2 x (1-x)~.
\end{equation}

For the spin-dependent case, the only contribution is the term proportional to 
$G_M^{(1)}$, which comes from the $A_1^{\rm Born}$ term (this is the only term that contributes to the Born (Zemach) piece of the hyperfine splitting). 
For the spin-independent case we only need $G_E^{(2)}$.

Eq.~(\ref{TBorn}) comes from diagrams (5) and (6) in Figs.~\ref{piL} and \ref{D1L} after properly subtracting the subdivergences. 

Following common practice we define the electromagnetic charge density as
\be
\rho_e(r)
\equiv
\int \frac{d^3k}{(2\pi)^3} e^{i{\bf k}\cdot {\bf r}}G_E(-{\bf k}^2)
\;.
\ee
The inverse of its Fourier transform allows us to obtain the even powers of the moments of the charge distribution of the proton, 
\be
\label{eq:GE}
G_E(-{\bf k}^2)=\sum_{n=0}^{\infty}\frac{(-1)^n}{(2n+1)!}{\bf k}^{2n}\int_0^{\infty}dr(4\pi)r^{2n}\rho_e(r)
=\sum_{n=0}^{\infty}\frac{(-1)^n}{(2n+1)!}{\bf k}^{2n}\langle r^{2n} \rangle
\,.
\ee
By Taylor expanding Eq. (\ref{GE2}) we obtain (for $k > 1$)
\bea
\label{r2kan}
&&
\langle r^{2k}\rangle=\frac{\mpi^{2-2 k}}{32 F_\pi^2 \pi ^2} \left(1+g_A^2 (3+2 k)\right) k(k-1) \Gamma(k-1)^2
\\
&+&\frac{ \mpi^{2-2k} }{36 F_\pi^2 \pi ^2 y^2}g_{\pi N \Delta}^2\left\{k\left(\frac{(3+2 k)}{1-k} y^2-6\right) \Gamma(k)^2
+ \ln(2)\frac{(-1)^{k+1} 4^{1-k} (3+2 k)(2 k)!}{(2 k-1)} y^{2k}\frac{\sqrt{1-y^2}}{\left(1-y^2\right)^k}\right\}\nn\\
&+&\frac{ \mpi^{2-2 k}}{18 F_\pi^2 \pi ^2}g_{\pi N \Delta}^2 y^{-4+2 k} \left(1-y^2\right)^{\frac{1}{2}-k} (k!)^2 \left\{  -3 \left(y^2-1\right)\binom{-1/2}{k-1} {}_3F_2\left(1,1,1-k;2,\frac{3}{2}-k;1-\frac{1}{y^2}\right)\right.\nn\\
&-&\left. 4 \left(y^2-1\right)\binom{-1/2}{k-2} {}_3F_2\left(1,1,2-k;2,\frac{5}{2}-k;1-\frac{1}{y^2}\right)-y^2 \ln\left(y^2\right)\left(4 \binom{-1/2}{k-1}+3 \binom{-1/2}{k}\right) \right\}\nn\\
&-& \frac{\mpi^{2-2 k}}{9 F_\pi^2 \pi ^2 }g_{\pi N \Delta}^2\frac{\left(y^2-1\right)^k}{y^2\left(1-y^2\right)^{\frac{1}{2}+k}} (k!)^2\sum _{r=1}^{\infty } \frac{ (2 r)!}{2^{2 r+1}r (r!)^2}y^{2 r}\left[\left(3+y^2\right) \binom{r}{k} {}_2F_1\left(-k,\frac{1}{2},1-k+r,\frac{y^2}{y^2-1}\right)\right.\nn\\
&-&\left.4 y^2 \binom{1+r}{k} {}_2F_1\left(-k,\frac{1}{2},2-k+r,\frac{y^2}{y^2-1}\right)\right]
\,,
\nn
\eea
%% %%%%%%%%%%%%%%%
where $y\equiv \frac{\mpi}{\Delta}$, and $\Gamma(n)$ is the Euler $\Gamma$ function.

The odd powers of the moments of the charge distribution of the proton are obtained (defined) through the relation:
\be
\label{rMoments}
\langle r^{2k+1}\rangle
=
\frac{\pi^{3/2}\Gamma(2+k)}{\Gamma(-1/2-k)}2^{4+2k}
\int \frac{d^3q}{(2\pi)^3}\frac{1}{{\bf q}^{2(2+k)}}
\left[
G_E(-{\bf q}^2)-\sum_{n=0}^k\frac{{\bf q}^{2n}}{n!} \left(\frac{d}{d{\bf q}^2}\right)^nG_E(-{\bf q}^2)\Big|_{{\bf q}^2=0}
\right]
\,.
\ee
An analytic expression of this quantity is relegated to Eq. (\ref{rZemach}). 
Note that, by using dimensional regularization, we can eliminate all the terms proportional to integer even powers of ${\bf q}^2$ in this expression. 
For $k>1$, this integral is dominated by the chiral result and can be approximated by
\be
\langle r^{2k+1}\rangle
\simeq
\frac{\pi^{3/2}\Gamma(2+k)}{\Gamma(-1/2-k)}2^{4+2k}
\int \frac{d^{D-1}q}{(2\pi)^{D-1}}\frac{1}{{\bf q}^{2(2+k)}}
G^{(2)}_E(-{\bf q}^2)
\,.
\ee

Finally, let us note that, by construction, both $T^{\mu\nu}_{\rm Born}$ and $T^{\mu\nu}_{\rm pol}$ 
 comply with current conservation. The separation 
(definition) of the Born and polarizability terms is in general ambiguous, see, for instance, the discussion in 
Refs. \cite{Scherer:1996ux,Fearing:1996gs}. 
In our case, as far as we give an explicit definition 
for $T^{\mu\nu}_{\rm Born}$, this ambiguity disappears. 
In what follows we consider the computation of $T^{\mu\nu}_{\rm pol}$. 

\subsection{Computation of $T^{\mu\nu}_{\rm pol}$}
\label{Sec:Tpol}
We split each $S^{\rm pol}_i/A^{\rm pol}_i$ in the following way:
\be
S_i^{\rm pol}=S_{i,\pi}^{\rm pol}+S_{i,\Delta}^{\rm pol}+S_{i,\pi\Delta}^{\rm pol}
\,,
\quad
A_i^{\rm pol}=A_{i,\pi}^{\rm pol}+A_{i,\Delta}^{\rm pol}+A_{i,\pi\Delta}^{\rm pol}
\,.
\ee 
$S_{i,\pi}^{\rm pol}$ and $A_{i,\pi}^{\rm pol}$ encode the contributions only due to pions. They are produced by the diagrams listed in 
Fig. \ref{piL}. 
Summing them up we can reconstruct the tensor structure of $T^{\mu\nu}$. 
In $D$ dimensions the structure functions read
\bea
S^{\rm pol}_{1,\pi}(q^2,\qv)
&=&
-\frac{g_A^2} {  F_\pi^2} M_p \left(\mpi^2 J_0'\left(0,\mpi^2\right)+J_0\left(0,\mpi^2\right)-J_0\left(\qv,\mpi^2\right)\right.\nn\\
&+&\left.4\intx \left\{(2x-1)J_2'\left(\qv x,\mt^2\right)-(1-x)\left(\mt^2+(q^2-2\qv^2)x^2\right)J_2''\left(\qv x,\mt^2\right)\right\}\right)\nn\\
&+&(\qv\rightarrow -\qv)
,\\
%\eea
%\bea
S^{\rm pol}_{2,\pi}(q^2,\qv)&=&\frac{g_A^2 }{ F_\pi^2 }M_p \mpi \frac{q^2}{  \qv^2}\left(J_0\left(0,\mpi^2\right)+\mpi^2 J_0'\left(0,\mpi^2\right)-J_0\left(\qv,\mpi^2\right)\right.\nn\\
&+&\left.\intx \left\{q^2\vq^2(1-2x)^2(1-x)x^2J_0''\left(\qv x,\mt^2\right)+2q^2(2x-1)xJ_0'\left(\qv x,\mt^2\right)\right.\right.\nn\\
&-&(1-x)\left(4(\mt^2-2\qv^2x^2)+q^2(4x^2+(2x-1)(1+6x+d(2x-1)))\right)J_2''\left(\qv x,\mt^2\right)\nn\\
&+&\left.\left.4(2x-1)J_2'\left(\qv x,\mt^2\right)\right\}\right)+(\qv\rightarrow -\qv) ,
\\
%\eea
%\bea
A^{\rm pol}_{1,\pi}(q^2,\qv)&=&-2\frac{g_A^2}{ F_\pi^2}M_p^2\intx\left\{\frac{1}{\qv}J_2'\left(\qv x,\mt^2\right)+\qv x^2 J_0'\left(\qv x,\mt^2\right)+x\mathcal D_\pi(\mt^2)\right\}\nn\\
&+&(\qv\rightarrow -\qv)
,
%\eea
%\begin{eqnarray}
\\
A^{\rm pol}_{2,\pi}(q^2,\qv)&=&\frac{g_A^2}{ F_\pi^2}M_p^3\intx x (2 x-1) J_0'\left(\qv x,\mt^2\right)-(\qv\rightarrow -\qv)
\,,
%\end{eqnarray}
\eea
where the loop functions $J_i$ have been defined in $D$-dimensions in \eq{Js}.

These structure functions reduce to the following expressions in $D=4$:
\bea
\label{S1pol}
{S}_{1,\pi}^{\rm pol}(q^2,\qv)&=&\frac{1}{\pi}\:\left(\frac{g_{A}}{2F_\pi}\right)^2\:M_p\:m_{\pi}\:\left\{\frac{3}{2}+\frac{m_{\pi}^2}{{\bf q}^2}-\left(1+\frac{m_{\pi}^2}{{\bf q}^2}\right)\:\sqrt{1-z}\right.\\
&-&\left.\frac{1}{2}\:\sqrt{\frac{m_{\pi}^2}{{\bf q}^2}}\:\left(2+\frac{q^2}{{\bf q}^2}\right)\:\mathcal{I}_{1}\,(m_{\pi}^2,q^{0},q^2)\right\}\nonumber
\,,
%\eea
%\bea
\\
\label{S2pol}
{S}_{2,\pi}^{\rm pol}(q^2,\qv)&=&\frac{1}{\pi}\:\left(\frac{g_{A}}{2F_\pi}\right)^2\:M_p\:m_{\pi}\:
\frac{q^2}{{\bf q}^2}
\left\{-\left(\frac{3}{2}+\left(\frac{1}{2}+\frac{m_{\pi}^{2}}{q^{2}}
+\frac{m_{\pi}^{2}}{\left(q^{0}\right)^{2}}\right)\frac{q^{2}}{{\bf q}^{2}}\right)\right.\nonumber\\
&-&\frac{\left(q^{0}\right)^{2}\:q^{2}}{4m_{\pi}^{2}{\bf q}^{2}+(q^{2})^{2}}\:\left(\frac{m_{\pi}^2}{{\bf q}^2}-\frac{{ q}^2}{2{\bf q}^2}\right)\\
&+&\frac{m_{\pi}^{2}}{{\bf q}^{2}}\left(2-\frac{{\bf q}^{2}}{ \left(q^{0}\right)^{2}}\left(1-z\right)+\frac{q^{2}\:\left(q^{0}\right)^{2}}{4m_{\pi}^{2}{\bf q}^{2}+(q^{2})^{2}}\right)\:\sqrt{1-z}\nonumber\\
&+&\left.\frac{1}{2}\sqrt{\frac{m_{\pi}^{2}}{{\bf q}^{2}}}\:\left(2+3\frac{q^{2}}{{\bf q}^{2}}+\frac{q^{2}}{m_{\pi}^{2}}\right)\:\mathcal{I}_{1}\,(m_{\pi}^2,q^{0},q^2)\right\}\nonumber
\,,
\\
%\eea
%\begin{eqnarray}
A_{1,\pi}^{\rm pol}(q^2,\qv)&=&-\frac{1}{2\pi^2}\frac{g_A^2}{ F_\pi^2}M_p^2\intx\frac{\sqrt{\mt^2}}{\qv}\left(\frac{\qv x}{\sqrt{\mt^2}}- \left(1-\frac{\qv^2 x^2}{\mt^2}\right)^{-1/2}\sin ^{-1}\left(\frac{\qv x}{\sqrt{\mt^2}}\right) \right),\nn\\
\\
%\end{eqnarray}
%\begin{eqnarray}
A_{2,\pi}^{\rm pol}(q^2,\qv)&=&-\frac{1}{4\pi^2}\frac{g_A^2}{ F_\pi^2}M_p^3\intx \frac{x (2 x-1)}{\sqrt{\mt^2}} \left(1-\frac{\qv^2 x^2}{\mt^2}\right)^{-1/2} \sin ^{-1}\left(\frac{\qv x}{\sqrt{\mt^2}}\right),
%\end{eqnarray}
\eea
where
\be
z=\frac{(q^{0})^2}{m_{\pi}^2},
\ee
and
\bea\label{eq:integraldef1}
\mathcal{I}_{1}\,(m_{\pi}^2,q^{0},q^2)
&=&\:\int_{0}^{1}dx\frac{1}{\sqrt{\frac{m_{\pi}^{2}}{{\bf q}^{2}}-\frac{q^{2}}{{\bf q}{2}}\:x-x^{2}}}
\\&=&
-\arctan\left(\frac{q^2}{2m_{\pi}|{\bf q}|}\right)
+\arctan\left(\frac{2{\bf q}^2+q^2}{2|{\bf q}|\sqrt{m_{\pi}^2-q_0^2}}\right)
\nn
\\
\nn
&=&i\ln \left(\frac{2i \mpi \sqrt{\vq^2}-q^2}{2 i\sqrt{\vq^2} \sqrt{\mpi^2-\qv^2} +q^2-2 \qv^2}\right)
\,.
\eea
For $D=4$ we can compare with previous results in the literature. 
$S^{\rm pol}_{1,\pi}$ and $S^{\rm pol}_{2,\pi}$ were originally computed in \cite{Nevado:2007dd}. We agree with those results, which were obtained with different methods, either by dispersion relations or through a diagrammatic computation assuming gauge invariance. In the case of real photons (for $q^2=0$ in the Coulomb gauge) we recover the results of \cite{Bernard:1992qa}. $S^{\rm pol}_{1,\pi}$ has also been checked in the limit $q_0 \rightarrow 0$ in Ref. \cite{Birse:2012eb}, 
and $S^{\rm pol}_{1/2,\pi}$ for all $q_0$ and $q^2$ in Ref. \cite{Alarcon:2013cba}. 

The spin-dependent structure functions, 
$A^{\rm pol}_{1,\pi}$ and $A^{\rm pol}_{2,\pi}$,  agree with the ones given in Eqs. (30) and (34) of \cite{Ji:1999mr}, up to a normalization factor. They follow from summing up all the contributions of the diagrams in \fig{piL} that have an antisymmetric contribution, i.e. diagrams (2), (4) and (5) of Fig. \ref{piL}.

We now move to contributions involving Delta particles. We first consider tree-level Delta mediated contributions. 
The corresponding diagram is pictured in \fig{DTL}, and the associated contributions read:
\bea
S_{1,\Delta}^{\rm pol}(q^2,\qv)&=&-\frac{4}{9}\frac{b_{1F}^2}{M_p^2}M_p\frac{  \Delta \vq^2}{\qv^2-\Delta ^2+i\eta},\label{S1TL}\\
S_{2,\Delta}^{\rm pol}(q^2,\qv)&=&\frac{4}{9}\frac{b_{1F}^2}{M_p^2}M_p\frac{\Delta  q^2}{\qv^2-\Delta ^2+i\eta},
\label{S2TL}\\
A_{1,\Delta}^{\rm pol}(q^2,\qv)&=&\frac{4 b_{1F}^2}{9 M_p^2}M_p^2\frac{  \qv^2}{ \qv^2-\Delta ^2+i\eta},
\label{A1TL}\\
A_{2,\Delta}^{\rm pol}(q^2,\qv)&=&-\frac{4 b_{1F}^2}{9 M_p^2}M_p^3\frac{ \qv}{\qv^2-\Delta ^2+i\eta}.
\label{A2TL}\eea
\eq{S1TL} agrees with  \cite{Nevado:2007dd} and, in the limit $\qv \rightarrow 0$, with the leading order expression 
of \cite{Birse:2012eb} up to normalization. \eq{S1TL} differs from the expression obtained in Ref.~\cite{Nevado:2007dd} using 
dispersion relations by a local term. For the spin-dependent terms we are in agreement with \cite{Ji:1999mr}.

The last set of diagrams that we consider are those with one internal chiral loop and virtual Delta particles. They are 
drawn in \fig{D1L} producing the following $D$-dimensional expressions for the structure functions: 
\bea
S^{\rm pol}_{1,\pi\Delta}(q^2,\qv)&=&-\frac{32}{3}\frac{D-2}{D-1}M_p \left(\frac{g_{\pi N \Delta }}{F_\pi}\right)^2 \left(\frac{1}{4} (D-1) J_2'\left(-\Delta ,\mpi^2\right)-\frac{1}{4} J_0\left(\qv-\Delta ,\mpi^2\right)\right.\nn\\
&-&\left.\intx\left\{(1-x) \left(-\Delta ^2+\mt^2+q^2 x^2+2 \qv x (\Delta -\qv x)\right) J_2''\left(\qv x-\Delta ,\mt^2\right)\right.\right.\nn\\
&+&\left.\left.\frac{\Delta}{D}(1-x)\left(\mt^2 \mathcal{D}_\pi''\left(\mt^2\right)+2  \mathcal{D}_\pi' \left(\mt^2\right)\right)+(2 x-1)J_2'\left(\qv x-\Delta ,\mt^2\right)\right\}\right)\nn\\
&+&(\qv\rightarrow -\qv),
\eea
\bea
S^{\rm pol}_{2,\pi\Delta}(q^2,\qv)&=&-\frac{8}{3}\frac{D-2}{D-1} M_p \frac{q^2}{\qv^2} \left(\frac{g_{\pi N\Delta }}{F_\pi}\right)^2\left(J_0\left(\qv-\Delta ,\mpi^2\right)-(D-1) J_2'\left(-\Delta ,\mpi^2\right)\right.\nn\\
&+&\left.\intx\frac{}{}\left\{(1-x) J_2''\left(\qv x-\Delta ,\mt^2\right) \left(q^2 \left(D (1-2 x)^2-4 x (1-4 x)-1\right)\right.\right.\right.\nn\\
&+&\left.\left.\left.4 \mt^2-4 \left(\Delta ^2+2 \qv x (\qv x-\Delta )\right)\right)+2 q^2 x (1-2 x) J_0'\left(\qv x-\Delta ,\mt^2\right)\right.\right.\nn\\
&+&\left.\left.q^2 \vq^2 (1-x) x^2 (2 x-1) (1-2 x) J_0''\left(\qv x-\Delta ,\mt^2\right)\right.\right.\nn\\
&-&\left.\left.4 (2 x-1) J_2'\left(\qv x-\Delta ,\mt^2\right)+\frac{4\Delta}{D}(1-x)\left(  \mt^2  \mathcal{ D}_\pi''\left(\mt^2\right)+2 \mathcal {D}_\pi'\left(\mt^2\right)\right)\right\}\right)\nn\\
&+&(\qv\rightarrow -\qv),
\eea
\bea
A^{\rm pol}_{1,\pi\Delta}(q^2,\qv)&=&-\left(\frac{g_{\pi N \Delta}}{F_\pi }\right)^2M_p^2\frac{16}{3(D-1)} \intx\left\{ x(\Delta+\qv x)J_0'\left(-\qv x-\Delta,\mt^2\right)\right.\nn\\
&-&\left.x\mathcal D_\pi'(\mt^2)\frac{1}{\qv}J_2'\left(-\qv x-\Delta,\mt^2\right)\right\}+(\qv\rightarrow -\qv),\nn\\
\\
A^{\rm pol}_{2,\pi\Delta}(q^2,\qv)&=&-\left(\frac{g_{\pi N \Delta}}{F_\pi }\right)^2M_p^3\frac{8 }{3(D-1)}\intx x(1-2x)J_0'\left(-\qv x-\Delta,\mt^2\right)\nn\\
&-&(\qv\rightarrow -\qv).
\eea

The results for $D=4$ dimensions are:
\bea
S^{\rm pol}_{1,\pi\Delta}(q^2,\qv)&=&-\frac{4 }{9 \pi^2}\frac{g_{\pi N \Delta}^2}{F_\pi^2} M_p \mpi\left[ 3 \mathcal{Z}\left(\frac{\Delta }{\mpi}\right)-\mathcal{Z}\left(\frac{\Delta-\qv}{\mpi}\right)-\mathcal{Z}\left(\frac{\Delta +\qv}{\mpi}\right)\right.\nn\\
&+&\left.\intx\left\{\frac{\Delta}{\mpi}  (5x-3) \ln \left(\frac{\mt^2}{\mpi^2}\right)+\sqrt{\frac{\mt^2}{\mpi^2}} \left(\left(5x-3+\frac{\vq^2 (1-x) x^2}{\mt^2-(\Delta +\qv x)^2}\right) \right.\right.\right.\nn\\
&&\left.\left.\left.\mathcal{Z}\left(\frac{\Delta +\qv x}{\sqrt{\mt^2}}\right)+\left(5x-3+\frac{\vq^2 (1-x) x^2}{\mt^2-(\Delta -\qv x)^2}\right) \mathcal{Z}\left(\frac{\Delta -\qv x}{\sqrt{\mt^2}}\right)\right)\right\}\right],
%\eea
%\bea
\\
S^{\rm pol}_{2,\pi\Delta}(q^2,\qv)&=&-\frac{4}{9 \pi ^2}\frac{g_{\pi N \Delta}^2}{F_\pi^2}M_p \mpi\frac{ q^2 }{\qv^2}\left[ -3 \mathcal{Z}\left(\frac{\Delta }{\mpi}\right)+\mathcal{Z}\left(\frac{\Delta-\qv}{\mpi}\right)+\mathcal{Z}\left(\frac{\Delta +\qv}{\mpi}\right)\right.\nn\\
&+&\left.\intx \left\{\frac{\Delta}{\mpi}(3x-5) \ln \left(\frac{\mt^2}{\mpi^2}\right)+\frac{1}{4}\sqrt{\frac{\mt^2}{\mpi^2}} \mathcal{Z}\left(\frac{\Delta +\qv x}{\sqrt{\mt^2}}\right)\left(\frac{}{}4 (3-5x)\right.\right.\right.\nn\\
&+&\left.\left. \left.\frac{(3-7x)(1-2x)^2q^2-4\vq^2x^2}{\mt^2-(\Delta +\qv x)^2}+\frac{q^2 \vq^2 (1-x) x^2 (1-2 x)^2}{\left(\mt^2-(\Delta +\qv x)^2\right)^2}\right)\right.\right.\nn\\
&+&\left.\left.\frac{1}{4}\sqrt{\frac{\mt^2}{\mpi^2}}\mathcal{Z}\left(\frac{\Delta -\qv x}{\sqrt{\mt^2}}\right)\left(\frac{}{}4 (3-5x)+\frac{(3-7x)(1-2x)^2q^2-4\vq^2x^2}{\mt^2-(\Delta -\qv x)^2}\right.\right.\right.\nn\\
&&\left.\left. \left.+\frac{q^2 \vq^2 (1-x) x^2 (1-2 x)^2}{\left(\mt^2-(\Delta -\qv x)^2\right)^2}\right)+\frac{q^2 \vq^2}{4\mt^2} (1-2 x)^2 (1-x) x^2\right. \right.\nn\\
&&\left.\left.\left(\frac{\Delta +\qv x}{\mpi \left(\mt^2-(\Delta +\qv x)^2\right)}+\frac{\Delta-\qv x }{\mpi \left(\mt^2-(\Delta -\qv x)^2\right)}\right)\right\}\right],
\eea
\bea
A^{\rm pol}_{1,\pi\Delta}(q^2,\qv)&=&\frac{2}{9\pi^2}\frac{g_{\pi N \Delta}^2}{F_\pi^2}M_p^2\left(1-\intx \sqrt{\mt^2}\left\{ \left(-\frac{1}{\qv}+\frac{ x (\Delta -\qv x)}{\mt^2-(\Delta -\qv x)^2}\right)\right.\right.\nn\\
&&\left.\left.\mathcal{Z}\left(\frac{\Delta -\qv x}{\sqrt{\mt^2}}\right)+\left(\frac{1}{\qv}+\frac{ x (\Delta +\qv x)}{\mt^2-(\Delta +\qv x)^2}\right)\mathcal{Z}\left(\frac{\Delta +\qv x}{\sqrt{\mt^2}}\right)\right\}\right),
%\eea
%\bea
\\
A^{\rm pol}_{2,\pi\Delta}(q^2,\qv)&=&\frac{1}{9 \pi ^2}\frac{g_{\pi N \Delta}^2}{F_\pi^2} M_p^3\intx\, x (1-2x)\sqrt{\mt^2}\left\{\frac{\mathcal{Z}\left(\frac{\Delta -\qv x}{\sqrt{\mt^2}}\right)}{\mt^2-(\Delta -\qv x)^2}-\frac{\mathcal{Z}\left(\frac{\Delta +\qv x}{\sqrt{\mt^2}}\right)}{\mt^2-(\Delta +\qv x)^2}\right\}.\nn\\
\eea
where we have defined $\mathcal Z$  as 
\beq
\mathcal Z(x)\equiv\sqrt{x^2-1} \ln \left(\sqrt{x^2-1}+x\right).\label{Z}
\eeq

The $D=4$ expressions for $S^{\rm pol}_{1,\pi\Delta}$ and $S^{\rm pol}_{2,\pi\Delta}$
agree with Eqs. (51) in \cite{Hemmert:1996rw} for the case of real photons, i.e. $q^2=0$ and in the Coulomb gauge.

Summing up all the contributions of the diagrams in \fig{D1L} which have an antisymmetric contribution, i.e. diagrams (2), (4) and (5), we get the spin-dependent part that
agrees with Eqs. (33) and (36) in \cite{Ji:1999mr}, up to a normalization factor.

In all expressions we use principal value prescriptions, the Dirac delta contributions associated to the propagators have gone into the Born term. Nevertheless, from the point of view of the effective theory this splitting between the polarizability and Born term is quite arbitrary.

\section{Matching HBET to NRQED: $c_3^{pl_i}$ (spin-independent)}
\label{Sec:c3}

The matching between HBET and NRQED can be performed in a generic
expansion in $1/M_p$, $1/m_\mu$ and $\al$.  We have two sort of loops:
chiral and electromagnetic.  The former are always associated to
$1/(4\pi F_0)^2$ factors, whereas the latter are always suppressed by
$\al$ factors. Any scale left to get the dimensions right scales with
$m_\pi$ or $\Delta$. In our case we are only concerned with obtaining the matching
coefficients of the lepton-baryon operators of NRQCD with
${\cal O}\left(\al^2\times\left(\frac{m_{l_i}}{m_{\pi}},\frac{m_{l_i}}{\Delta}\right)\right)$
accuracy. 

In what follows, we will assume that we are doing the 
matching to NRQED($\mu$). Therefore, we keep the whole dependence on 
$m_{l_i}/m_\pi$. The NRQED($e$) case can then be derived by 
expanding $m_e$ versus $m_\pi$. 

At ${\cal O}(\al^2)$, the contribution to $c^{pl_i}_{3}$ (see Fig. \ref{fig:c3}) 
from matching HBET to NRQED can be written in a compact way in 
terms of the structure functions of the forward virtual-photon Compton 
tensor. It reads \cite{Bernabeu:1973uf}
\bea
\label{c3}
c_{3}^{pl_i}
&=&
- e^4 M_pm_{l_i}\int {d^4k_E \over (2\pi)^4}{1 \over k_E^4}{1 \over
k_E^4+4m_{l_i}^2k_{0,E}^2 }
%\\
\nn
%&&
%\times
\left\{
(3k_{0,E}^2+{\bf k}^2)S_1(ik_{0,E},-k_E^2)-{\bf k}^2S_2(ik_{0,E},-k_E^2)
\right\}
\\
&&
+{\cal O}(\al^3)
\,.
\eea
This result keeps the complete dependence on $m_{l_i}$ and is valid both for NRQED($\mu$) and NRQED($e$). This contribution 
is usually organized in the following way
\be
\label{c3split}
c_{3}^{pl_i}=
c_{3,\rm R}^{pl_i}
+ c_{3,\rm point-like}^{pl_i}
+ c_{3,\rm Born}^{pl_i}+ c_{3,\rm pol}^{pl_i} +{\cal O}(\al^3)
\,.
\ee
$c_{3,\rm R}^{pl_i}$ is suppressed by 
an extra factor $m_{l_i}/M_p$, i.e. $c_{3,\rm R}^{pl_i} 
\sim \al^2 m_{l_i}/M_p$. This
goes beyond the aimed accuracy of our calculation and so we neglect 
$c_{3,\rm R}^{pl_i}$.

The second term in Eq.~(\ref{c3split}) corresponds to Eq.~(\ref{c3}) assuming the proton 
to be point-like. With the precision needed it reads in the $\MS$ scheme (see \cite{Pineda:1998kn})\footnote{In this expression we have computed the loop with the proton being relativistic to follow common practice. 
Nevertheless, this assumes that one can consider the proton to be point-like at the scales of the proton mass. 
To stick to an standard EFT approach one should consider the proton to be NR. Then one would obtain
\be
\label{c3pointlike}
 c_{3,\rm point-like}^{pl_i}
=\al^2{M_p \over m_{l_i}}\left(\ln{m_{l_i}^2 \over \nu^2}+{1 \over 3}\right)
\,.
\ee
The difference between both results is of the order of $c_{3,\rm R}$, and gets absorbed into this coefficient 
(which we do not know anyhow). Therefore, the value of $c_{3}^{pl_i}$, will be the same no matter the prescription used. In practice there could be some difference due to truncation, but always of the order of the error of our computation.
}
\be
c_{3,\rm point-like}^{pl_i}(\nu)\equiv
\frac{M_p}{m_{l_i}} {\al^2 \over M_p^2-m^2_{l_i}}
\left\{M_p^2\left(  \ln{m^2_{l_i} \over  \nu^2}
                   + {1 \over 3} \right)
       -
       m^2_{l_i}\left(  \ln{M^2_p \over  \nu^2}
                   + {1 \over 3} \right)
\right\}
\,,\ee

\subsection{$c_{3,\rm Born}$ and Zemach moments}

\begin{figure}[h]
\makebox[2.0cm]{\phantom b}
\epsfxsize=9truecm \epsfbox{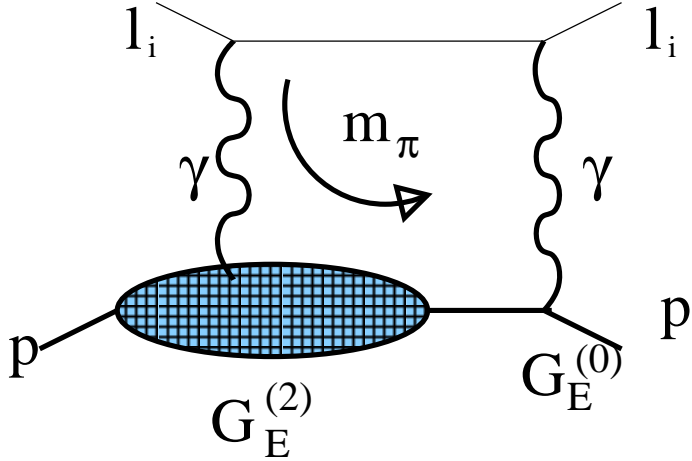}
\caption {{\it Symbolic representation (plus permutations) 
of the Zemach correction in Eq. (\ref{c3Zemach}).}}
\label{figzemach}
\end{figure}

The third term in Eq. (\ref{c3split}) is generated by the spin-independent 
Born contribution to $T^{\mu\nu}$ in Eq. (\ref{TBorn}). We symbolically picture it in 
Fig. \ref{figzemach}. At leading order in the NR expansion it reads\footnote{In Ref.~\cite{Pineda:2004mx} we named this object $c_{3,Zemach}^{pl_i}$.}
\be
\label{c3Zemach}
c_{3,\rm Born}^{pl_i}= 
4(4\pi\alpha)^2M_p^2m_{l_i}\int {d^{D-1}q \over (2\pi )^{D-1}}
{1 \over {\bf q}^6}G_E^{(0)}G_E^{(2)}(-{\bf q}^2)
\,.
\ee
Note again that this result holds for both NRQED($e$) and NRQED($\mu$). In other 
words, the exact dependence on $m_{l_i}$ is kept (at leading order in the NR expansion). 
The linear dependence in the lepton mass 
 makes this contribution much smaller for the case of hydrogen.
$G_E^{(0)}=1$. We take the 
expression for $G_E^{(2)}$ from Eq. (\ref{GE2}). 
The use of effective field theories and dimensional regularization 
is a strong simplification, which we have already used when writing Eq.~(\ref{c3Zemach}). This guarantees that only low energy modes contribute 
to the integral, and that we only need the non-analytic behavior of $G_E^{(2)}$ 
in $q^2$ around $m_{\pi}$ and $\Delta$. In other words, 
even though some point-like contributions are still encoded in $G_E^{(2)}$, they do do not contribute to the integral. 
The analytical behavior in $q^2$ produces scaleless integrals, which are zero 
in dimensional regularization. This is a reflection of the factorization of the 
different scales. Therefore, we do not need to introduce the point-like interactions to 
regulate the infrared divergences of the integrals at zero momentum, as it is done if trying to compute this object 
directly from the experimental data. We will come back to this issue when we discuss the Zemach moments.

The computation of $c_{3,\rm Born}^{pl_i}$ was made in Ref.~\cite{Pineda:2004mx}. Here we give a simplified expression: 
\bea
\label{c3Born}
c_{3,\rm Born}^{pl_i}&=& 
{2}(\pi\al)^2 
\left({M_p \over 4\pi F_0 }\right)^2
{m_{l_i} \over m_\pi}
\left\{
{3 \over 4}g_A^2+{1 \over 8}+\frac{32}{9}\pi g_{\pi N\Delta}^2\frac{m_{\pi}^2}{\Delta^2-m_{\pi}^2}
\right.
\\
&&
\left.
+{2 \over \pi}g_{\pi N\Delta}^2{m_\pi \over \Delta}
\sum_{r=0}^{\infty}{(-1)^r\Gamma(-3/2) \over \Gamma(r+1)\Gamma(-3/2-r)}
\left\{
B_{6+2r}-{2(r+2) \over 3+2r}B_{4+2r} 
\right\}\left({m_\pi \over \Delta}\right)^{2r}
\right\}
\,,
\nn
\eea
where the first line is due to scales of ${\cal O}(m_\pi)$ and the terms proportional to $B_n$ are due to scales of
${\cal O}(\Delta)$, where (this corrects Eq. (61) of Ref.~\cite{Pineda:2004mx})
\bea
B_n&\equiv& \int_0^\infty \,dt
{t^{2-n} \over 1-t^2}
\times
 \,\left\{
\begin{array}{ll}
 \displaystyle{
\sqrt{1-t^2}\ln{\left[{1\over t}+\sqrt{{1 \over t^2}-1}\right]}}
 &{\rm if} \, \,t<1\;
\\
 \displaystyle{
-\sqrt{t^2-1}\arccos[{1 \over t}]}
 &{\rm if} \,\, t > 1\;
\end{array}
\right.
\\
\nn
&=&
-\frac{\sqrt{\pi } \left(H_{\frac{1}{2}-\frac{n}{2}}-H_{1-\frac{n}{2}}\right) \Gamma \left(\frac{3}{2}-\frac{n}{2}\right)}{4 \Gamma \left(2-\frac{n}{2}\right)}   
	 \\
	 \nn
	 &&
	 + 2^{1 - n}\,\pi \,{{\Gamma}( n-2) \over \Gamma^2(\frac{n}{2})}\,
   {{}_3F_2}( \frac{1}{2},\frac{ n-2}{2},\frac{ n-1}{2}; 
    \frac{n}{2},\frac{n}{2}; 1) 
   \\
   \nn
   &&
  +\frac{2^{\frac{5}{2}-\frac{n}{2}} \,
   _3F_2\left(\frac{3}{2}-\frac{n}{2},\frac{3}{2}-\frac{n}{2},\frac{n}{2}+\frac{1}{2};\frac{5}{2}-\frac{n}{
   2},\frac{5}{2}-\frac{n}{2};\frac{1}{2}\right)}{(n-3)^2}
   \\
   \nn
   &&
   -\frac{2^{\frac{3}{2}-\frac{n}{2}} \,
   _3F_2\left(\frac{5}{2}-\frac{n}{2},\frac{5}{2}-\frac{n}{2},\frac{n}{2}+\frac{1}{2};\frac{7}{2}-\frac{n}{
   2},\frac{7}{2}-\frac{n}{2};\frac{1}{2}\right)}{(n-5)^2}
   \\
   \nn
   &&
   +\frac{\pi ^{3/2} \sec \left(\frac{\pi 
   n}{2}\right) \left((n-2) H_{1-n}+(2-n) H_{\frac{1}{2}-\frac{n}{2}}+n (-\ln (2))-1
   +\ln(4)\right)}{(n-2) \Gamma \left(2-\frac{n}{2}\right) \Gamma \left(\frac{n-1}{2}\right)}
   	\,,
\eea
and $H_n$ is the $n$ harmonic number.

Eq.~(\ref{c3Born}) encapsulates all the non-analytic dependence in the light quark masses  
and in the splitting between the nucleon and the Delta mass (proportional to powers of $1/N_c$ in the large $N_c$ limit)  
of $c_{3,\rm Born}^{pl_i}$. 
This expression is the leading contribution to the Zemach term in 
the chiral counting (supplemented with a large $N_c$ counting). This 
is a model independent result.  Other contributions to the Zemach term 
are suppressed in the chiral counting.

$c_{3,\rm Born}^{pl_i}$ can be related with (one of) the Zemach moments:
\be
\langle r^m \rangle_{(2)}
\equiv
\int d^3 r r^m \int d^3 z \rho_e(|{\bf z}-{\bf r}|)\rho_e(z).
\ee
The Zemach moments can be determined in a similar way as the moments of the charge distribution of the proton.
For even powers we have the relation\footnote{Note that comparison with Eq.~(\ref{eq:GE}) gives algebraic relations between the 
even charge, $\langle r^{2n} \rangle$, and Zemach, $\langle r^{2n} \rangle_{(2)}$, moments.}
\be
G_E^2(-{\bf k}^2)
=\sum_{n=0}^{\infty}\frac{(-1)^n}{(2n+1)!}{\bf k}^{2n}\langle r^{2n} \rangle_{(2)}.
\ee
The odd powers are obtained (defined) through the relation:
\be
\langle r^{2k+1}\rangle_{(2)}
=
\frac{\pi^{3/2}\Gamma[2+k]}{\Gamma[-1/2-k]}2^{4+2k}
\int \frac{d^3q}{(2\pi)^3}\frac{1}{{\bf q}^{2(2+k)}}
\left[
G^2_E(-{\bf q}^2)-\sum_{n=0}^k\frac{{\bf q}^{2n}}{n!} \left(\frac{d}{d{\bf q}^2}\right)^nG^2_E(-{\bf q}^2)\Big|_{{\bf q}^2=0}
\right]
\,.
\ee
Again, using dimensional regularization, we can eliminate all the terms proportional to integer even powers of ${\bf q}^2$ in this expression. 
For $k \geq 1$ this integral is dominated by the chiral result and can be approximated by
\be
\langle r^{2k+1}\rangle_{(2)}
\simeq
2\times\frac{\pi^{3/2}\Gamma[2+k]}{\Gamma[-1/2-k]}2^{4+2k}
\int \frac{d^{D-1}q}{(2\pi)^{D-1}}\frac{1}{{\bf q}^{2(2+k)}}
G^{(2)}_E(-{\bf q}^2)
\simeq
2\langle r^{2k+1}\rangle
\,.
\ee
It is possible to get an analytic result for these integrals. We obtain ($y \equiv \frac{m_{\pi}}{\Delta}$) 
\bea
\label{rZemach}
\langle r^{2k+1}\rangle_{(2)}
&\simeq&
2\langle r^{2k+1}\rangle
\simeq
2\Gamma[3/2+k]
\frac{m_{\pi}^{1-2k}}{(4\pi F_0)^2}
\left\{
\Gamma[3/2+k]
\frac{2+4g_A^2(2+k)}{3+4(k^2-1)}
\right.
\\
\nn
&+&\frac{4}{9}g^2_{\pi N\Delta}\frac{\pi(k+2)(-1)^{k+1}}{\Gamma[5/2-k]}y^2
\;{}_2F_1( \frac{3}{2},1; 
    \frac{5}{2}-k; y^2)
\\
\nn
&+&
\left.
\frac{32}{3}g^2_{\pi N\Delta}y^{2k-1}
\sum_{r=0}^{\infty}\frac{y^{2r}}{r!}\frac{(-1)^{r}}{\Gamma[-1/2-k-r]}
\left[
B_{2k+2r+4}-\frac{r+ {4\over 3}k+{2\over 3}}{\frac{1}{2}+k+r}B_{2k+2r+2} 
\right]
\right\}
\,.
\eea

%%%%%%%%%%%%%%%%%% Tabla masa%%%%%%%%%%%%%%%%%%%%%%%%%%%%%%%%%%%%%%%5
\begin{table}[h]
\addtolength{\arraycolsep}{0.2cm}
$$
\begin{array}{|l||c|c|c|c|c|c|}
\hline
  & \langle r^3 \rangle & \langle r^4 \rangle & \langle r^5 \rangle & \langle r^6 \rangle & \langle r^7 \rangle &  \langle r^3 \rangle_{(2)}
\\ \hline\hline
\pi & 
0.4980
%0.497998 
& 
0.6877
%0.687655 
& 
1.619
%1.61909  
& 
5.203
%5.20311 
& 
20.92
%20.9241 
& 
0.9960
%0.995997
\\
\pi\&\Delta & 
0.4071
%0.40709 
& 
0.6228
%0.62277 
& 
1.522
%1.52155  
& 
4.978
%4.97769 
& 
20.22
%20.2229  
& 
0.8142
%0.81418
\\ \hline
\cite{Janssens:1965kd}& 0.7706  & 1.083 & 1.775 & 3.325  &7.006  & 2.023 \\
%{\;Q<m_\rho} &0.4125   & & 0.9179  &    &3.5517   &1.25078 \\
%{\;Q<M_p} &0.4758   & &1.0240  &    &3.9910   &1.36626 \\  
\cite{Kelly:2004hm}  & 
0.9838
%0.98379 
& 
1.621
%1.62132 
& 
3.209
%3.2089 
& 7.440 & 
19.69
%19.6876 
& 
2.526
%2.5260 
\\
\cite{Distler:2010zq}  & 1.16(4) & 2.59(19)(04) & 8.0(1.2)(1.0) & 29.8(7.6)(12.6) & ---& 2.85(8) \\
%{\;Q<m_\rho}  &0.741027  &  & 5.6985 &  &   &1.92676 \\ 
%{\;Q<M_p}  &0.793234  &  & 6.06737 &  &   &2.06929 \\ 
     \hline
\end{array}
$$
\caption{{\it Values of $\langle r^n \rangle$ in fermi units. The first two rows give the prediction from the effective theory: the first 
row for the effective theory with only pions and the second for the theory with pions and Deltas. The third row 
corresponds to the standard dipole fit of Ref.~\cite{Janssens:1965kd} with $\langle r^2 \rangle=0.6581$ fm$^3$. 
The fourth and fifth rows correspond to different parameterizations of experimental data \cite{Kelly:2004hm,Distler:2010zq}, 
with the latest fit being the more recent analysis based on Mainz data. For completeness, 
we also quote $\langle r^3 \rangle_{(2)}=2.71$ fm$^3$ from Ref.~\cite{Friar:2005jz}.}}
\label{tab:rnZemach}
\end{table}
%%%%%%%%%%%%%%Tabla masa%%%%%%%%%%%%%%%%%%%%%%%%%%%%%%%%%%

In Table \ref{tab:rnZemach} we give our predictions for some selected charge and Zemach moments,\footnote{Note that 
$\langle r^{2k+1}\rangle_{(2)}
\simeq
2\langle r^{2k+1}\rangle$ with the precision of our computation.} both in the effective theory with 
only pions and in the effective theory with pions and Deltas. The even powers are obtained by direct numerical Taylor expansion of 
Eq.~(\ref{GE2}), or using the analytic formulas in Eq.~(\ref{r2kan}). The odd powers are obtained from Eq. (\ref{rZemach}). We have also numerically checked the values of 
$\langle r^{2k+1} \rangle$ directly using Eq.~(\ref{rMoments}). 
In order to estimate the error of the charge/Zemach moments and the other quantities we compute in this paper 
we proceed as follows. 
We count $m_{\pi} \sim \sqrt{\lQ m_q}$ and $\Delta \sim \frac{\lQ}{N_c}$. We then have the double expansion 
$\frac{m_{\pi}}{\lQ} \sim \sqrt{\frac{m_q}{\lQ}}$ and $\frac{\Delta}{\lQ} \sim \frac{1}{N_c}$. We still have to determine the 
relative size between $m_{\pi}$ and $\Delta$. We observe that  $m_{\pi}/\Delta \sim N_c \sqrt{\frac{m_q}{\lQ}} \sim 1/2$.
Therefore, we associate a 50\% uncertainty to the pure chiral computation. For all Zemach moments we observe good 
convergence, with the contribution due to the Delta being much smaller than the pure chiral result, and well inside the 50\%
uncertainty. Leaving aside the Delta, the splitting with the next resonances suggest a mass gap of order $\lQ \sim$ 500-770 
MeV depending on whether one considers the Roper resonance or the $\rho$. For practical purposes, 
we also count $m_K \sim \sqrt{\lQ m_s} \sim 500$ MeV of order $\lQ$.  
Therefore, we assign $\frac{m_{\pi}}{\lQ} \sim 1/3$ and $\frac{\Delta}{\lQ} \sim 1/2$, as the uncertainties of the pure chiral 
and the Delta-related contribution respectively. We add these errors linearly for the final error. This gives the expected size of the uncomputed corrections but numerical factors may change the real size of the correction. In particular, huge discrepancies with these estimates may signal the failure of HB$\chi$PT for obtaining some of the observables considered in this paper. 

The chiral prediction is expected to 
give the dominant contribution of $\langle r^n \rangle$ for $n \geq 3$. For $n=2$ it could also give the leading chiral log. 
For smaller $n$ the chiral corrections are subleading. Note that for all $n \geq 3$,  
these expressions give the leading (non-analytic) dependence 
in the light quark mass as well as in $1/N_c$. This is a valuable information for eventual 
lattice simulations of these quantities where one can tune 
these parameters. In Table \ref{tab:rnZemach} we also compare with the standard dipole ansatz  \cite{Janssens:1965kd}, 
and with different determinations using experimental data of the electric Sachs form factor fitted to more sophisticated functions 
\cite{Kelly:2004hm,Distler:2010zq}.\footnote{The agreement with \cite{Kelly:2004hm} for $n=7$ is accidental. We have checked that the
growth with $n$ is different with respect the chiral prediction.} The latest fit claims to be the more accurate. Nevertheless, we observe large differences, bigger than the errors. This is specially worrisome for large $n$, since the chiral prediction is expected to 
give the dominant contribution of $\langle r^n \rangle$ for $n \geq 3$. In this respect, we believe that the chiral result may help to shape the appropriated fit function and, thus, to discriminate between different options, as well as to assess uncertainties. The impact of 
choosing different fit functions can be fully appreciated, for instance, in the different values of the electromagnetic proton radius 
obtained in Ref. \cite{Bernauer:2010wm} versus Refs. \cite{Lorenz:2012tm,Lorenz:2014vha} from direct fits to the $ep$ scattering data. Such values differ by around 3 standard deviations. On the other hand, even if on general grounds one may expect 
the charge/Zemach moments will be more and more sensitive to the chiral region for $n \rightarrow \infty$, large fractions of the experimental numbers are determined by the subtraction terms included to render these objects finite (for odd powers of $n$). 
We stop the discussion here but the reason for such large discrepancies should be further investigated. 

As we have already mentioned, $c_{3,\rm Born}^{pl_i}$ can be related with (one of) the Zemach moments:
\be
c_{3,\rm Born}^{pl_i}=\frac{\pi}{3}\al^2M_p^2m_{l_i}\langle r^3 \rangle_{(2)}
\,,
\qquad
\langle r^3 \rangle_{(2)}= \frac{48}{\pi}
\int_0^{\infty} {d Q \over Q^4}
\left(
G_E^2(-Q^2)-1+\frac{Q^2}{3}\langle r^2\rangle
\right)\,.
\ee
Note again that the terms proportional to "1" and $r^2$ vanish in dimensional regularization. 
We can now obtain  ($m_r=m_{\mu}M_p/(m_{\mu}+M_p)$)
\be
\Delta E_{\mathrm{Born}}=\frac{c_{3,\rm Born}^{pl_{\mu}}}{M_p^2}\frac{1}{\pi}\left(\frac{m_r\al}{2}\right)^3
\;,
\ee
the Born contribution to $\Delta E_{\mathrm{TPE}}$, 
from the effective field theory. We quote our results in Table \ref{Table:EnZemach}. The pure chiral result was already 
obtained in Ref.~\cite{Pineda:2004mx}.
The $\pi\&\Delta$ result corrects the evaluation made in that reference due to the error in its Eq. (61). Note that the new result is much more convergent, since the correction associated to the Delta is much smaller. On the other hand, our result is now much more different with respect to standard values obtained from dispersion relations. We quote two of them in Table \ref{Table:EnZemach}. One may wonder whether such difference is due to relativistic corrections. 
An estimate of the relativistic effects can be obtained from the analysis made in Ref.~\cite{pachucki1}, which, however,
is based on dipole form factors parameterizations. The difference between the relativistic and NR expression was found to be small ($\sim 3 \mu$eV). It should be checked whether this feature holds with different parameterizations. If so, the difference seems to be mainly due to the computation of the Zemach correction (see Table \ref{tab:rnZemach} and the discussion above). Therefore, as stated above, 
the reason for such large discrepancies should be investigated. In the mean time we will stick to our model independent prediction from the effective theory.
 
\begin{table}[h]
\centering
$$
%\footnotesize
\begin{array}{|c|ccc|ccc|}                          
\hline
\mu {\rm eV}    &  {\rm DR}& \cite{Pachucki:1999zza}   &   \cite{Carlson:2011zd}  & {\rm HBET} & \cite{Pineda:2004mx} (\pi) & (\pi\&\Delta)
\\
\hline
\Delta E_{\mathrm{Born}}             & & 23.2(1.0)     &    24.7(1.6)     && 10.1(5.1) & 8.3(4.3)
                                                                                                \\                                                                                            
\hline
\end{array}
$$
\caption{\it Predictions for the Born contribution to the $n=2$ Lamb shift. The first two entries correspond to dispersion relations. The 
last two entries are the predictions of HBET:  The 3rd entry is the prediction of HBET at leading order (only pions) and the last entry is the 
prediction of HBET at leading and next-to-leading order (pions and Deltas). 
\label{Table:EnZemach}}
\end{table}
 
\subsection{Matching HBET to NRQED: $c^{p l_i}_{3,\rm pol}$}

Finally, we consider the polarizability correction. It is obtained from Eq.~(\ref{c3}) but subtracting the Born term to the structure functions of the virtual-photon Compton tensor. The expressions at ${\cal O}(p^3)$ in HB$\chi$PT can be found in Sec.~\ref{Sec:Tpol}. The final expression reads

\bea
\label{Eq:c3pol}
c^{pl_i}_{3,\rm pol}&=&-e^4 M_p^2 \frac{m_\mu}{m_\pi}\left(\frac{g_A}{F_\pi} \right)^2 \mathcal I_2^\pi
-e^4 b_{1F}^2 \frac{m_\mu}{\Delta}\frac{4}{9} \mathcal I_2^\Delta-e^4 M_p^2 \frac{m_\mu}{\Delta}\frac{8}{3}\left(\frac{g_{\pi N \Delta}}{F_\pi} \right)^2 \mathcal I_2^{\Delta\pi},
\eea

where
\bea
\mathcal I_2^{i}&=&\int \frac{d^3k}{(2\pi)^3}\frac{1}{(1+{\bf k}^2)^4}\int_0^\infty \frac{d\omega}{\pi}\frac{1}{\omega}\frac{1}{\omega^2+4\hat m_i \frac{1}{(1+{\bf k}^2)^2}}\left\{\left(2+(1+{\bf k}^2)^2\right)A_E^i \left(\omega,{\bf k}^2\right)\right.\nn\\
&+&\left.(1+{\bf k}^2)^2{\bf k}^2\omega^2B_E^i \left(\omega,{\bf k}^2\right)\right\}.
\eea

For the case of only pions we have $\hat m_\pi=m_\mu/m_\pi$ and
\bea
A_E^\pi \left(\omega,{\bf k}^2\right)&=&-\frac{1}{4\pi}\left[-\frac{3}{2}+\sqrt{1+\omega^2}+\int_0^1 dx \frac{1-x}{\sqrt{1+x^2 \omega^2+x(1-x)\omega^2{\bf k}^2}}
\right]
\,,
\\
B_E^\pi \left(\omega,{\bf k}^2\right)&=&\frac{1}{8\pi}\int_0^1 dx \left[\frac{1-2x}{\sqrt{1+x^2 \omega^2+x(1-x)\omega^2{\bf k}^2}}-\frac{1}{2} \frac{(1-x)(1-2x)^2}{\left(1+x^2 \omega^2+x(1-x)\omega^2{\bf k}^2\right)^{\frac{3}{2}}}\right]\nn
\,.
\\
\eea

For the case of Delta at tree level we have $\hat m_\Delta=m_\mu/\Delta$ and 
\bea
A_E^\Delta \left(\omega,{\bf k}^2\right)&=&\frac{1}{\pi^2}\frac{\omega^2 {\bf k^2}}{\omega^2+1},\\
B_E^\Delta \left(\omega,{\bf k}^2\right)&=&-\frac{1}{\pi^2}\frac{1}{\omega^2+1}.
\eea

For the case of loops including the Delta we have $\hat m_{\Delta\pi}=m_\mu/\Delta$ and 
\bea
A_E^{\Delta \pi}\left(\omega,{\bf k}^2\right)&=&-\frac{1}{12 \pi ^2}\intx \left\{3 \sqrt{1-t^2} \ln\left(\frac{1+\sqrt{1-t^2}}{t}\right)\right.\nn\\
&+&\left.2\sqrt{-t^2-(i+\omega )^2} \left(\ln(t)-\ln\left(1-i \omega +\sqrt{-t^2-(i+\omega )^2}\right)\right)\right.\nn\\
&-&\left.(3-5 x) \ln\left(1+\frac{\left(1+{\bf k}^2\right) (1-x) x \omega ^2}{t^2}\right)\right.\nn\\
&+&\left.2\frac{\left(t^2-1\right) (3-5 x)+2 i x (3-5 x) \omega +x \left(3-5 x+3{\bf k}^2(1-x) (1-2 x)\right) \omega ^2}{\sqrt{1-t^2+x \omega  \left(-2 i+\left(-1+{\bf k}^2 (-1+x)\right) \omega \right)}}\right.\nn\\
&&\left.\ln\left(\frac{1-i x \omega +\sqrt{1-t^2+x \omega  \left(-2 i+\left(-1+{\bf k}^2 (-1+x)\right) \omega \right)}}{\sqrt{t^2-\left(1+{\bf k}^2\right) (-1+x) x \omega ^2}}\right)\right\}+(\omega \rightarrow -\omega),\nn\\
\\
B_E^{\Delta \pi}\left(\omega,{\bf k}^2\right)&=&\frac{1}{24 \pi ^2}\intx(1-2 x)^2 \nn\\
&&\left\{\frac{ {\bf k}^2 (1-x) x^2\omega ^2 (1+i \omega  x)}{\left(t^2-\left(1+{\bf k}^2\right) (-1+x) x \omega ^2\right) \left(1-t^2+x \omega  \left(2 i+\left(-1+{\bf k}^2 (-1+x)\right) \omega \right)\right)}\right.\nn\\
&-&\left.\frac{-3+t^2 (3-7 x)+x \left(7+2 i (-3+7 x) \omega +\left(3-7 x+3 {\bf k}^2 (1-x) (1-2 x)\right) \omega ^2\right)}{\left(1-t^2+x \omega  \left(2 i+\left(-1+{\bf k}^2 (-1+x)\right) \omega \right)\right)^{3/2}}\right.\nn\\
&&\left.\ln\left(\frac{1+i x \omega +\sqrt{1-t^2+x \omega  \left(2 i+\left(-1-{\bf k}^2 (1-x)\right) \omega \right)}}{\sqrt{t^2-\left(1+{\bf k}^2\right) (-1+x) x \omega ^2}}\right)\right\}+(\omega \rightarrow -\omega),\nn\\
\eea
where $t=\mpi/\Delta$. 
Note that the imaginary part of these expressions comes only from the Wick rotation of $k_0$ and will vanish upon integration.

The pure pion contribution was already found in Ref.~\cite{Nevado:2007dd}. 
Our full prediction for the polarizability term including the Delta effects reads
\be
\label{Epol}
\Delta E_{\mathrm{pol}}=\frac{c_{3,\rm pol}^{pl_{\mu}}}{M_p^2}\frac{1}{\pi}\left(\frac{m_r\al}{2}\right)^3
=18.51(\pi{\rm -loop})-1.58(\Delta{\rm -tree})+9.25(\pi\Delta{\rm -loop})=26.2(10.0) \mu{\rm eV}
\,.
\ee
In Table \ref{Table:Epol} we compare our determination with previous results. Most of them are obtained by a combination of 
dispersion relations plus some modeling of the subtraction term that we discuss below. The analysis of Ref.~\cite{Alarcon:2013cba} has a different status. 
In this reference the polarizability correction was computed using $ {\rm B}\chi{\rm PT}$ with only pions.  Such computation treats the baryon relativistically. The result incorporates some subleading effects, which are sometimes used to give an estimate of higher order effects in $ {\rm HB}\chi{\rm PT}$. Nevertheless, the computation also assumes that a theory with only baryons and pions is appropriate at the proton mass scale. This should be taken with due caution. Still, it would be desirable to have a deeper theoretical understanding of this difference, which may signal that relativistic corrections are important for the polarizability correction. In any case, the  $ {\rm B}\chi{\rm PT}$ computation differs of our chiral result by around 50\% (this means around 1.5 times the error we use for the chiral contribution, once the Delta is incorporated in the calculation), which we consider reasonable. 

It is also worth discussing the LEX approximation used in Ref.~\cite{Alarcon:2013cba}. This approximation consists in setting $q_0=0$ 
everywhere except in the denominator in Eq.~(\ref{c3}). For the pure chiral result, this approximation works remarkable well (18.51%45
 (exact) vs 17.85 (LEX)). Nevertheless, such success does
not survive the incorporation of the $\Delta$ particle. For the Delta tree-level contribution we find (-1.58%6
 (exact) vs 0 (LEX)). The real 
problem appears from the ${\cal O}(p^3)$ pion-Delta result. For such contribution there are $1/q_0$ singularities in the tensor that only cancel if the complete 
expression is used. Doing the LEX approximation leads to divergent expressions. Even more worrisome is the fact that, at present, there are 
no theoretical justification for using the LEX approximation for the integral in Eq. (\ref{c3}). It is not correct to assume that the photon energy that appears in the integral, $q_0$, corresponds to the energy in the atomic system. It rather reflects virtual fluctuations of order of the pion and muon mass (as well as of the $\Delta$ scale). Since those particles are relativistic at those scales it is theoretically incorrect, a priori, to neglect $q_0$.  In any case, on the light of the good agreement for the pure chiral 
case, it would be interesting to see whether one could find a theoretical justification for such behavior.
 
\begin{table}[h]
\centering
$$
\begin{array}{|c|ccccc|c|cc|}                          
\hline
(\mu {\rm eV})  & {\rm DR+Model}& \cite{Pachucki:1999zza}   &  \cite{Martynenko:2005rc}  & \cite{Carlson:2011zd}  &  \cite{Gorchtein:2013yga}   &  {\rm B}\chi{\rm PT} \cite{Alarcon:2013cba}  (\pi) & {\rm HBET} \cite{Nevado:2007dd} (\pi) & \cite{Peset:2014yha}(\pi\&\Delta)\\
\hline
\Delta E_{\mathrm{pol}} &  & 12(2)       &    11.5     &   7.4  (2.4)   &   15.3(5.6)    & 8.2 (^{+1.2}_{-2.5}) & 18.5(9.3) & 26.2(10.0)
\\
\hline
\end{array}
$$
\caption{\it Predictions for the polarizability contribution to the $n=2$ Lamb shift. The first four entries use dispersion relations for the inelastic term and different modeling
functions for the subtraction term. The number of the fourth entry has been taken from \cite{Alarcon:2013cba}. 
The 5th entry is the prediction obtained using B$\chi$PT. The last two entries are the predictions of HBET discussed in this paper. 
The 6th entry is the prediction at leading order (only pions) and the last entry is the 
prediction at leading and next-to-leading order (pions and Deltas). 
\label{Table:Epol}}
\end{table}

\begin{figure}[h]
\makebox[2.0cm]{\phantom b}
\epsfxsize=9truecm \epsfbox{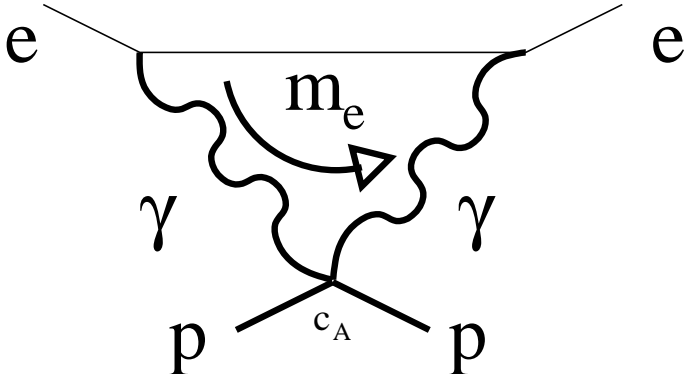}
\caption {{\it Diagram contributing to the
polarizability correction with $\ln m_e$ accuracy. The matching coefficients 
of the proton can be $c_{A_1}$ or $c_{A_2}$, or, in other words, the 
proton polarizabilities.}}
\label{figpolhyd}
\end{figure}

It is also interesting to consider the limit $m_{l_i} \ll m_{\pi}$, which is relevant for the 
hydrogen atom. In this limit 
Eq. (\ref{Eq:c3pol}) approximates, with logarithmic accuracy, to  
\be
c_{3,\rm pol}^{pl_i}=-\al M_p^2m_{l_i}\left[5\alpha_{E}^{(p)}-\beta_{M}^{(p)}\right]
\ln({m_{l_i}})
\,.
\ee
\bea
\label{c3log}
&& c_{3,\rm pol}^{pl_i}=-{2 \over 9}\al^2 
{m_{l_i} \over \Delta}
b_{1,F}^2\ln{\Delta \over m_{l_i}}+{49 \over 12}\pi\al^2 g_A^2{m_{l_i}\over m_{\pi}} 
{M_p^2 \over 
(4\pi F_0)^2}
\ln({m_{\pi} \over m_{l_i}})
\\
\nn
&&
+{8 \over 27}\al^2 g_{\pi N\Delta}^2{m_{l_i}\over \sqrt{\Delta^2-m_{\pi}^2}} {M_p^2 
\over 
(4\pi F_0)^2}
\left({45 \Delta \over \sqrt{\Delta^2-m_{\pi}^2}}
+{4\Delta^2-49m_{\pi}^2 \over \Delta^2-m_{\pi}^2}\ln [{\cal R}\left(m_{\pi}^2\right)]\right)\ln({m_{\pi} \over 
m_{l_i}})
\,.
\eea
These logs can be obtained by computing the ultraviolet behavior of the diagram 
in Fig. \ref{figpolhyd}. This contribution is proportional to $c_{A_1}$ and $c_{A_2}$ or, in other words,
the polarizabilities of the proton (see \cite{Friar:1997tr,KS}).
For the pure pion cloud, the polarizabilities were computed in Ref. 
\cite{Bernard:1992qa}. 
The contribution due to the $\Delta$ can be found in Ref. \cite{Hemmert:1999pz}.
The scale in the logarithm is compensated by the next scale of the problem, 
which can be $m_{\pi}$ or $\Delta$. For contributions which are only due 
to the $\Delta$ or pions, the scale is unambiguous. In the case where pions 
and $\Delta$ are both present in the loop we will choose the pion mass (the 
difference being beyond the logarithmic accuracy). It is known that the pure chiral prediction of $\al^{(p)}_E$ 
 and $\beta_{M}^{(p)}$ nicely agrees with the experimental values. This agreement deteriorates after the 
 inclusion of the Delta effects, specially for $\beta_{M}^{(p)}$. Nevertheless, this object is comparatively small, and even more so 
 for $5\alpha_{E}^{(p)}-\beta_{M}^{(p)}$, the combination that appears in the logarithmic approximation. Whereas the experimental number reads $5\alpha_{E}^{(p)}-\beta_{M}^{(p)} \simeq 54 \times 10^{-4}$ fm$^{3}$ \cite{Beringer:1900zz},  
 the pure chiral result gives $(5\alpha_{E}^{(p)}-\beta_{M}^{(p)})(\pi) \simeq 60 \times 10^{-4}$ fm$^{3}$, and after the
 inclusion of the Delta we obtain $(5\alpha_{E}^{(p)}-\beta_{M}^{(p)})(\pi\&\Delta) \simeq 73 \times 10^{-4}$ fm$^{3}$. Again the 
 inclusion of the Delta deteriorates the agreement but the difference is of the order of one sigma according to our error 
analysis. We take this as an indication that effective field theory result will not be very far off from the real number for the case of muonic hydrogen and that, 
maybe, the pure chiral result compares better with experiment than after the inclusion of the Delta. Nevertheless, we will not make any assumption in this respect and stick to the complete prediction of the effective theory.
 
It is also customary to split the polarizability term (note that the Born term has already been subtracted from it) in what is called the inelastic and subtraction term:
\bea
c_{3,\rm sub}^{pl_i}
&=&
- e^4 M_pm_{l_i}\int {d^4k_E \over (2\pi)^4}{1 \over k_E^4}{1 \over
k_E^4+4m_{l_i}^2k_{0,E}^2 }
\nn
(3k_{0,E}^2+{\bf k}^2)S_1(0,-k_E^2)
\\
&=&
-\frac{\al^2 M_p}{2m_{l_i}}
\int_0^{\infty}\frac{d Q^2}{Q^2}
\left\{
1+\left(1-\frac{Q^2}{2 m_{l_i}^2}\right)\left(\sqrt{\frac{4m^2_{l_i}}{Q^2}+1}-1\right)
\right\}
S_1(0,-Q^2)
\eea
\bea
c_{3,\rm inel}^{pl_i}
&=&
- e^4 M_pm_{l_i}\int {d^4k_E \over (2\pi)^4}{1 \over k_E^4}{1 \over
k_E^4+4m_{l_i}^2k_{0,E}^2 }
\\
\nn
&&
\times
\left\{
(3k_{0,E}^2+{\bf k}^2)(S_1(ik_{0,E},-k_E^2)-S_1(0,-k_E^2))-{\bf k}^2S_2(ik_{0,E},-k_E^2)
\right\}
\eea
It is argued that the inelastic term does not require further subtractions and can be obtained through dispersion relations. On the other hand, the 
subtraction term cannot be directly obtained from experiment. 
This fact has been used in Ref.~\cite{Hill:2011wy} to emphasize that the polarizability term is affected by huge theoretical uncertainties. 
In this paper, we can avoid making any assumption about the dispersion relation properties of these quantities. This is possible within the 
framework of effective field theories. In this setup the splitting between the inelastic and subtraction terms is unmotivated, and to some 
extent artificial (as it was the splitting between the Born and polarizability term). Let us elaborate on this point and see what effective field theories have to say in this respect. The main problem comes, as it has already been 
pointed out in Ref.~\cite{Alarcon:2013cba}, from the diagram in Fig. \ref{DTL}.  This diagram yields a finite (an small) 
contribution to $c_{3,\rm pol}^{pl_i}$ 
(and therefore to the energy shift, see Eq. (\ref{Epol})). 
Nevertheless, when splitted into $c_{3,\rm inel}^{pl_i}$ and $c_{3,\rm sub}^{pl_i}$, each term diverges in the following way
\be
\delta c_{3,\rm sub}^{pl_i} \sim -\delta c_{3,\rm inel}^{pl_i} \simeq -\frac{4}{3}\al^2\frac{m_{l_i}}{\Delta}b_{1,F}^2\ln (\nu/m_{l_i})
\,.
\ee
If we set the ultraviolet cutoff to the $\rho$ mass, $\nu=m_{\rho}$, the energy shift of each term is one order of magnitude bigger $\sim -11.37$ $\mu$eV than the exact result for the sum. Obviously such contribution is fictitious and may alter the value of the individual terms. On the other hand, it is possible to perform this splitting for the case of the pion and pion-Delta loop. We obtain the following:
\be
\Delta E^{(\mathrm{sub})}(\pi{\rm -loop})=-1.62 \; \mu {\rm eV}\; ; \qquad \Delta E^{(\mathrm{sub})}(\pi\Delta{\rm -loop})=-1.23 \; \mu {\rm eV}.
\ee
They are of the same magnitude. Their size is barely one order of magnitude smaller than the total polarizability term. 
For the case of the pion loop it is possible to obtain analytic expressions in the limit $m_{\mu}=m_{\pi}$, which is a rather good approximation:
\be
\Delta E^{(\mathrm{sub})}(\pi{\rm -loop})\Bigg|_{m_{\pi}=m_{\mu}}
=-\frac{g_A^2 \alpha^5 m_r^3}{64 \pi^2 F_\pi^2}\frac{m_\mu}{m_\pi}\left(-1+3G-2\ln 2\right) = -1.40 
\; \mu {\rm eV},
\ee
where $G\simeq 0.9160$ is the Catalan's constant.
For these quantities the LEX approximation works quite well, both for the pion and the pion-Delta loop case. We find\footnote{For 
$m_{\mu}=m_{\pi}$ an analytic expression can be found for the pion-loop case \cite{Alarcon:2013cba}:
\be
\Delta E_{\rm LEX}^{(\mathrm{sub})}(\pi{\rm -loop})\Bigg|_{m_{\pi}=m_{\mu}}
=-\frac{g_A^2 \alpha^5 m_r^3}{64 \pi^2 F_\pi^2}\frac{m_\mu}{m_\pi}\left(\frac{1}{2}-G+\ln 2\right) = -1.08 
\; \mu {\rm eV}.
\ee}
\be
\Delta E_{\rm LEX}^{(\mathrm{sub})}(\pi{\rm -loop})=-1.23
%-1.22858 
\; \mu {\rm eV}\; ; 
\qquad \Delta E_{\rm LEX}^{(\mathrm{sub})}(\pi\Delta{\rm -loop})=-0.91
%0.907292
\; \mu {\rm eV},
\ee
which is again asking for a theoretical explanation of this relatively good agreement. 

For comparison we show different values 
obtained for the subtraction and inelastic term obtained in the literature in Table~\ref{Table:Sub}. 

\begin{table}[htb]
\centering
$$
\begin{array}{|c|ccccc|c|}                          
\hline
(\mu {\rm eV})  &   \cite{Pachucki:1999zza}   &  \cite{Martynenko:2005rc}  & \cite{Carlson:2011zd}   & \cite{Birse:2012eb}  &  \cite{Gorchtein:2013yga}                 &  \cite{Alarcon:2013cba}  \\
\hline
\Delta E^{(\mathrm{sub})} &  -1.8  &  -2.3  &  -5.3(1.9)  & -4.2(1.0) & 2.3 (4.6)^{(1)}                                & $3.0$  \\
\Delta E^{(\mathrm{inel})}            &   13.9                                                          &      13.8                                                    &  12.7(5)                               & ---   & 13.0(6)                                       & 5.2  \\ 
\hline
\end{array}
$$
\caption{\it Values for the subtraction and inelastic terms that one can find in the literature. $^{(1)}$This number is the adjusted value of Ref.~\cite{Gorchtein:2013yga}, given in \cite{Alarcon:2013cba}. 
\label{Table:Sub}}
\end{table}

We now combine the contribution from the Born and polarizability term and summarize our final results for 
$\Delta E_{\rm TPE}$:
\be
\Delta E_{\rm TPE}=\Delta E_{\rm Born}+\Delta E_{\rm pol}
=
28.59(\pi)+5.86(\pi\&\Delta)=34.4(12.5) \mu {\rm eV}
\,.
\ee
We would like to emphasize that this result is a pure prediction of the effective theory.
It is also the most precise expression that can be obtained in a model independent way, since ${\cal O}(m_{\mu}\alpha^5\frac{m_{\mu}^3}{\lQ^3})$ effects are not controlled by the chiral theory and would require new counterterms.
Our number is only marginally bigger than  $\Delta E_{\rm TPE}=33(2) \mu$eV \cite{Birse:2012eb}. 
This number is the one used in 
Ref.~\cite{Antognini:1900ns} for its determination of the proton radius. It is obtained as the sum of 
the elastic and inelastic terms from Ref.~\cite{Carlson:2011zd} and the subtraction term from 
Ref.~\cite{Birse:2012eb}. Note that this evaluation is model dependent. Even though the 
low energy behavior of the forward virtual Compton tensor was computed to ${\cal O}(p^4)$, 
this does not reflect in an improved determination of the polarizability correction, since 
an effective dipole form factor is used, not only at the $\rho$ mass scale, but also at the chiral 
scale. This problem also introduces a model dependence in its error estimate. 
Other existing determinations \cite{Pachucki:1999zza,Martynenko:2005rc,Carlson:2011zd} 
yield quite similar numbers but suffer from the same systematic uncertainties. In this respect our 
calculation is model independent and have completely different systematics. The fact that we obtain similar numbers 
is comforting for the reliability of the proton radius determinations obtained in Refs.~\cite{Antognini:1900ns,Peset:2014yha}. 
On the other hand, one should not forget that the individual contributions are quite different, and the reasons for that should be further 
investigated. Yet it is quite remarkable that the total sum gives such similar numbers.

\section{Matching HBET to NRQED: $c_4^{pl_i}$ (spin-dependent)}
\label{Sec:c4}

We proceed in the same way as in the spin-independent case. We will assume that we are doing the 
matching to NRQED($\mu$). Therefore, we keep the whole dependence on 
$m_{l_i}/m_\pi$. The NRQED($e$) case can then be derived by 
expanding $m_e$ versus $m_\pi$. We match 
 HBET and NRQED order by order  in a generic
expansion in $1/M_p$, $1/m_\mu$ and $\al$.  We have two sort of loops:
chiral and electromagnetic.  The former are always associated to
$1/(4\pi F_0)^2$ factors, whereas the latter are always suppressed by
$\al$ factors. Any scale left to get the dimensions right scales with
$m_\pi$ or $\Delta$. 

At ${\cal O}(\al^2)$, the contribution to $c^{pl_i}_{4}$ (see Fig. \ref{fig:c3}) 
from matching HBET to NRQED can be written in a compact way in 
terms of the structure functions of the forward virtual-photon Compton 
tensor. In Euclidean space it reads 
\bea
\nn
c_4^{pl_i}&=&{e^4 \over 3}\int {d^Dk \over (2\pi)^D}{1 \over k_E^2}{1 \over
k^4_E+4m_{l_i}^2k_{0,E}^2 }
\left\{
A_1(ik_{0,E},-k_E^2)(k_{0,E}^2+2k_E^2)+i3k_E^2{k_{0,E} \over M_p}A_2(ik_{0,E},-k_E^2)
\right\}
\\
&&
+{\cal O}(\al^3)
\,,
\label{c4}
\eea
consistent with the expressions obtained long ago in Ref. \cite{DS}. 
This result keeps the complete dependence on $m_{l_i}$ and is valid both for NRQED($\mu$) and NRQED($e$), 
i.e. for hydrogen and muonic hydrogen. 
Similarly to the spin-independent case, this contribution can be organized in the following way
\be
\label{c4split}
c_{4}^{pl_i}=
c_{4,\rm R}^{pl_i}
+ c_{4,\rm point-like}^{pl_i}
+ c_{4,\rm Born}^{pl_i}+ c_{4,\rm pol}^{pl_i} +{\cal O}(\al^3)
\,.
\ee
Within the effective field theory framework the contribution from energies of ${\cal O}(m_\rho)$
or higher in Eq. (\ref{c4}) are encoded in $c_{4,\rm R}^{pl_i}\simeq
c_{4,\rm R}^{p}$ (analogously to $c_3$). The other terms (associated to energies of ${\cal O}(m_\pi)$) were computed with 
${\cal O}(\al^2\times(\ln m_q, \ln \Delta, \ln m_{l_i}))$
accuracy in 
Ref.~\cite{Pineda:2002as}. We quote them here for ease of reference\footnote{In Ref.~\cite{Pineda:2002as} 
$c_{4,\rm Born}^{pl_i}$ was named $\delta c_{4,Zemach}^{pl_i}$, as  Eq.~(\ref{Zemachc4}) corresponds to the 
Zemach expression \cite{Zemach}, the leading order in the NR expansion of the Born term. The point-like contribution diverges
irrespectively of doing the computation in a relativistic or NR way (see 
the discussion in Ref.~\cite{Pineda:2002as}). Here we only quote the NR expression, which is more natural from the effective field 
theory point of view, as it avoids any assumption about the behavior of the theory at the proton mass scale.} ($c_F^{(p)}=1+\kappa_p$): 
\bea
\label{pointlike}
c_{4,\rm point-like}^{pl_i}&=&\left(1-\frac{ \kappa^2_p}{4}\right)\alpha^2\ln{m_{l_i}^2\over \nu^2} 
\,, 
%\ee
%\bea
\\
\label{Zemachc4}
c_{4,\rm Born}^{pl_i}
&\simeq&
(4\pi\alpha)^2M_p{2 \over 3}\int {d^{D-1}k \over (2\pi )^{D-1}}
{1 \over {\bf k}^4}G_E^{(0)}G_M^{(1)}
\\
&\simeq&
{M_p^2 \over (4\pi F_0)^2}\alpha^2{2 \over 3}\pi^2
\left[
g_A^2\ln{m_\pi^2 \over \nu^2}
+
{4 \over 9}g_{\pi N \Delta}^2
\ln{\Delta^2 \over \nu^2}
\right]
\,,
%\eea
%\be
\\
\label{polpiN}
c_{4,pol}^{pl_i}
&=&
{M_p^2 \over (4\pi F_0)^2}{\alpha^2 \over \pi}
{8\over 3}
\left(\frac{7\pi}{8}-\frac{\pi^3}{12}\right)
\left[
g_{A}^2\ln{m_\pi^2 \over \nu^2}
-\frac{8}{9}g_{\pi N\Delta}^2
\ln{\Delta^2 \over \nu^2}
\right]
+
{b_{1,F}^2 \over 18}\alpha^2\ln{\Delta^2 \over \nu^2}  
\,.
\eea
Summing up the three terms one has
\bea
\label{Zemachud}
c_{4}^{pl_i}
&\simeq&
\left(1-{\mu_p^2 \over 4}\right)\alpha^2\ln{m_{l_i}^2\over \nu^2}
+{b_{1,F}^2 \over 18}\alpha^2\ln{\Delta^2 \over \nu^2}  
 +
{m_p^2 \over (4\pi F_0)^2}\alpha^2{2 \over 3}
\left({2 \over 3}+{7 \over 2\pi^2}\right)\pi^2g_A^2\ln{m_\pi^2 \over \nu^2}
\nn
\\&&
+
{m_p^2 \over (4\pi F_0)^2}\alpha^2{8 \over 27}
\left({5 \over 3}-{7 \over \pi^2}\right)\pi^2g_{\pi N \Delta}^2
\ln{\Delta^2 \over \nu^2}
\\
&\stackrel{(N_c\rightarrow\infty)}{\simeq}&
\alpha^2\ln{m_{l}^2\over \nu^2}
 +
{m_p^2 \over (4\pi F_0)^2}\alpha^2
\pi^2g_A^2\ln{m_\pi^2 \over \nu^2}
\,.
\eea

Parametrically, the three contributions, Eqs.~(\ref{pointlike}), (\ref{Zemachc4}) and (\ref{polpiN}), are of the same order. 
Nevertheless, the polarizability and the point-like term are much smaller. 
This is consistent with the fact that the polarizability correction seems to be small \cite{DeRafael:1971mc,Faustov:2001pn,Carlson:2008ke}, if determined through dispersion relations. As already discussed in Ref.~\cite{Pineda:2002as},
the effective field theory computation gives a double explanation to this fact. On the one hand, this is due to the smallness of the numerical coefficient of the polarizability term, but there also seems to be some large $N_c$ 
rationale behind. Since $g_{\pi N\Delta}=3/(2\sqrt{2})g_A$ in the large $N_c$ limit, the polarizability term 
vanishes (see \cite{Ji:1999mr}) except for the tree-level-like Delta contribution (the last term in Eq.~(\ref{polpiN})). 
Nevertheless, the latter also vanishes against the $\kappa_p$-dependent point-like contribution 
(which effectively becomes the result of a point-like particle) in the large $N_c$ limit, 
since  $b_1^F=3/(2\sqrt{2})\kappa_V$ and $\kappa_p=\kappa_V/2$ \cite{KarlPaton}. 
Note also that the point-like term and the tree-level-like Delta contribution are suppressed by $1/\pi$ factors with respect the 
Born contribution.  

This discussion also illustrates that splitting the total contribution into different terms 
may introduce spurious effects that vanish in the total sum. We have also seen a 
similar thing but in a different context  for the case of the spin-independent computation. 

Our computation allows us to relate $c_4^{pl_{\mu}}$ and $c_4^{pl_{e}}$ in a model independent way. Since $c_{4,\rm R}^{pl_i}\simeq c_{4,\rm R}^{p}$ up to terms of ${\cal O}(\al^2 m_{l_i}/\lQ)$, 
we can obtain the following relation 
\be
\label{c4mue}
c_4^{pl_{\mu}}=c_4^{pl_{e}}+
\left[c_{4,\rm point-like}^{pl_{\mu}}-c_{4,\rm point-like}^{pl_{e}}\right]+ \left[c_{4,\rm pol}^{pl_{\mu}}
- c_{4,\rm pol}^{pl_{e}}\right] 
+{\cal O}(\al^3,\al^2 m_{\mu}/\lQ)
\,.
\ee
Note that we have already used the fact that $c_{4,\rm Born}^{pl_i}$ cancels in the difference, as it is independent of the lepton mass. 
The experimental and theoretical results discussed before suggest that $c_{4,\rm Born}^{pl_i}$ 
is the leading contribution to the Wilson coefficient. Therefore, such contribution can be obtained from $c_4^{pl_{e}}$, which can be determined from the hyperfine splitting of hydrogen. 
In Ref.~\cite{Pineda:2002as} it was estimated to be $c_4^{pl_{e}}\simeq -48\al^2$. 
By considering 
differences in Eq. (\ref{c4mue}) the ultraviolet behavior gets regulated and the logarithmic divergences vanish. 
This makes these contributions to be very small an negligible compared with the uncertainties. 
For the point-like contribution we obtain
\be
c_{4,\rm point-like}^{pl_{\mu}}-c_{4,\rm point-like}^{pl_{e}}=
\left(1-\frac{ \kappa^2_p}{4}\right)\alpha^2\ln{m_{\mu}^2\over m_e^2}
\simeq 2.09 \al^2
\,,
\ee
and for the polarizability we obtain (note that this term vanishes in the large $N_c$ limit, except for the tree-level-like contribution)
\be
c_{4,\rm pol}^{pl_{\mu}}
- c_{4,\rm pol}^{pl_{e}}
=
0.17\al^2(\pi)+0.07\al^2(\Delta)+0.008\al^2(\pi\&\Delta)=0.24\al^2
\,.
\ee
Overall we obtain $c_4^{pl_{\mu}}\simeq -46\al^2$. The bulk of this contribution is expected to come from the Born term, which in turn is related to the Zemach magnetic radius,
 \be
\langle r_Z \rangle=-\frac{4}{\pi}\int_0^{\infty}\frac{dQ}{Q^2}
\left[G_E(Q^2)G_M(Q^2)-1\right]
\ee
 by the following relation
\be
\langle r_Z \rangle=-\frac{3}{4\pi}\frac{1}{\al^2M_p}c_{4,\rm Born}^{pl_i}
\simeq
-{\pi \over 2}{M_p \over (4\pi F_0)^2}
\left[
g_{A}^2\ln{m_\pi^2 \over \nu^2}
+\frac{4}{9}g_{\pi N\Delta}^2
\ln{\Delta^2 \over \nu^2}
\right]
\stackrel{(\nu=m_{\rho})}{=}
1.35 \;{\rm fm}
\,.
\ee 
The chiral log result compares well ($\sim 30\%$) with existing predictions ($\sim 1.04$-$1.08$ fm) from hydrogen hyperfine \cite{Dupays:2003zz,Volotka:2004zu}, 
from dispersion relations \cite{Friar:2005jz,Distler:2010zq}, or from the muonic hydrogen hyperfine \cite{Antognini:1900ns}. 
Note that in the case of the determinations of $\langle r_Z \rangle$ from the hyperfine splitting (either from hydrogen or muonic hydrogen) 
one needs to control the relativistic hadronic affects associated to the Born term as well as the polarizability correction. On the 
other hand, if we are only interested in the hyperfine splitting it may make more sense to consider $c_4^{pl_i}$ as a whole. 
We relegate a more detailed discussion to future work.  
 
\section{Conclusions}

We have computed the spin-dependent and spin-independent structure functions of the forward virtual-photon Compton tensor 
of the proton at ${\cal O}(p^3)$ in HB$\chi$PT including the Delta particle. We have given D-dimensional expressions 
too. Those are relevant for future higher order loop computations. We have compared our results with previous computations. 
The $D=4$ expressions for the spin-dependent structure functions were computed in~\cite{Ji:1999mr}. We agree with their results. 
The $D=4$ expressions for the pure chiral (without Delta contributions) spin-independent structure functions were computed in~\cite{Nevado:2007dd}.
We agree with their results too. The Delta-associated contributions to the spin-independent structure functions are new. 
We also profit to present all these results obtained throughout the years in a unified form.

We have used these results to determine the leading chiral and large $N_c$ structure of $c^{pl_i}_3$ and $c^{pl_i}_4$, or, 
in other words, to determine their non-analytic dependence on $m_q$ and $N_c$. The fact that we have full control over  the quark mass dependence makes our result very useful for eventual lattice determinations of these quantities. By fine tunning the mass in simulations we can identify the 
results obtained in this paper and up to which mass the chiral is good approximation. One could also vary $N_c$ to check the theory.

These Wilson coefficients appear in the hyperfine splitting (spin-dependent) and Lamb shift (spin-independent) in hydrogen and muonic hydrogen. $c^{pl_i}_3$, 
the relevant Wilson coefficient for the Lamb shift, is chiral enhanced. Therefore, the ${\cal O}(p^3)$ chiral result is a pure prediction of the effective theory, 
which we use to determine 
\be
\label{ETPEfinal}
\Delta E_{\rm TPE}=
28.6(\pi)+6.1(\pi\&\Delta)=34.4(12.5) \mu {\rm eV}
\,,
\ee
the energy shift associated to the (hadronic) two-photon exchange of the Lamb shift. 
These results have been used in the recent determination of the muonic hydrogen Lamb shift and the proton radius performed in Ref.~\cite{Peset:2014yha}. We would like to emphasize that Eq. (\ref{ETPEfinal}) is the most precise expression that can be obtained in a model independent way, since ${\cal O}(m_{\mu}\alpha^5\frac{m_{\mu}^3}{\lQ^3})$ effects are not controlled by the chiral theory and would require new counterterms. Our final number is quite similar to previous estimates existing in the literature. Nevertheless, those computations require the splitting of the two-photon contribution into different terms. Some of them 
are then computed using different dispersion relations, whereas one last term requires modeling its $Q^2$ dependence.
In contrast, we have used the same method for all computations contributing to our result, yielding a parameter-free prediction.
On the other hand one should not forget that the individual contributions are quite different, and the reasons for that should be further 
investigated. In this respect we have discussed what the effective theory has to say about the separation into Born, polarizability, inelastic and subtraction term. The Born contribution is related with the Zemach moments. In this paper we have also given the prediction of the 
effective theory for some charge, $\langle r^{n} \rangle$, and Zemach, $\langle r^{n} \rangle_{(2)}$ moments.  Finally, we have also 
discussed the chiral dependence of the spin-dependent four-fermion Wilson coefficient, $c_4^{pl_i}$, 
and obtained the relation between $c_4^{p e}$ and $c_4^{p\mu}$ given by the effective theory.
 
\medskip

{\bf Acknowledgments} \\ 
We acknowledge discussions with F.~J.~Llanes-Estrada and V. Pascalutsa.
This work was supported by the Spanish grants FPA2010-16963 and FPA2011-25948, and the Catalan grant SGR2014-1450.
\appendix

%\label{app:formulas}

%%%%%%%%%%%%%%%%%%%%%%%%%%%%%%%%%%5
%\begin{appendices}
%\section{
\Appendix{Constants and parameters}
In our computations we have used the following values:
\bea
m_\mu&=&105.6583715 \MeV \\
m_\pi&=&139.57018 \MeV \\
M_p&=& 938.272046 \MeV \\
\Delta&=& 293.728 \MeV \\
\alpha&=&1/137.035999679\\
g_A&=&1.25\\
g_{\pi N \Delta}&=&1.05\\
F_\pi&=&92.5 \MeV\\
b_{1F}&=&3.86
\eea

The values of the masses and the fine structure constant come from the PDG database~\cite{Beringer:1900zz}. The values of the effective theory parameters correspond to the NR limit. 

\section{ Master Integrals}

We follow the notation of \cite{Bernard:1995dp,Hemmert:1996rw} and assume a negative infinitesimal imaginary part for all the propagators.
\bea
\frac{1}{i}\int \frac{d^Dl}{(2\pi)^D}\frac{\left\{1,l_\mu,l_\mu l_\nu,l_\mu l_\nu l_\alpha,l_\mu l_\nu l_\alpha l_\beta \right\}
}{(v\cdot l-q_0-i\eta)(m^2-l^2-i\eta)}&=&\{J_0(q_0,m), v_\mu J_1(q_0,m), g_{\mu\nu}J_2(q_0,m)+v_\mu v_\nu J_3(q_0,m),\nn\\
&&(g_{\mu\nu}v_\alpha+g_{\mu\alpha}v_\nu
+g_{\nu\alpha})v_\mu J_4(q_0,m)+v_\mu v_\nu v_\alpha J_5(q_0,m),\nn\\
&&(g_{\mu\nu}g_{\alpha\beta}+g_{\mu\alpha}g_{\nu\beta}
+g_{\nu\alpha}g_{\mu\beta})J_6(q_0,m)+(g_{\mu\nu}v_\alpha v_\beta\nn\\
&&+g_{\mu\alpha}v_\nu v_\beta)+g_{\mu\beta}v_\nu v_\alpha)+g_{\nu\alpha}v_\mu v_\beta)+g_{\alpha\beta}v_\mu v_\nu)J_7(q_0,m)\nn\\
&&+...\},\label{Js}
\eea

where in $D$-dimensions:
\bea
\mathcal{D}_\pi(m)&=&m^{D-2}(4\pi)^{-D/2}\Gamma\left(1-\frac{D}{2}\right)\\
J_0(\qv,m_\pi)&=&\frac{2}{(4\pi)^{D/2}}\Gamma\left(2-\frac{D}{2}\right)\int_{-\qv}^\infty dy \frac{1}{(m^2-\qv^2+y^2)^{2-D/2}}
\eea
and in $D=4-\epsilon$:
\bea
\mathcal{D}_\pi(m)&=&\frac{m^2}{16\pi^2}\tL\label{Dpi}+\mathcal O(\epsilon),\\
\tL&=&\mu^{2\epsilon}\left(\frac{1}{\epsilon}+(\gamma_E-1-\ln 4\pi)\right)+\ln\left(\frac{m^2}{\mu^2}\right)\label{tL}.
\eea
For the function $J_0$ we get when $\vert\omega\vert<m$
\bea
J_0(\qv,m_\pi)&=&\frac{\qv}{8 \pi ^2}(1-\tL)-\frac{1}{4\pi^2}\sqrt{\mpi^2-\qv^2}\cos^{-1}\frac{-\qv}{\mpi}+\mathcal O(\epsilon)\label{J0p},
\eea
and for the case where $\omega <-m$ we get the analytically continued function
\bea
J_0(q_0,m)&=&\frac{\qv}{8 \pi ^2}(1-\tL)+\frac{\sqrt{q_0^2-m^2} }{4 \pi ^2}\ln \left(\frac{\sqrt{q_0^2-m^2}-q_0}{m}\right)+\mathcal O(\epsilon)\label{J0}.
\eea
All the other functions are related to \eq{J0}/\eq{J0p} and \eq{Dpi}  by:
\bea
J_1(q_0,m)&=&q_0J_0(q_0,m)+\mathcal{D}_\pi(m),\\
J_2(q_0,m)&=&\frac{1}{D-1}((m^2-q_0^2)J_0(q_0,m)-q_0\mathcal{D}_\pi(m)),\\
J_3(q_0,m)&=&q_0J_1(q_0,m)-J_2(q_0,m), \\
J_4(q_0,m)&=&q_0J_2(q_0,m)+\frac{m^2}{D}\mathcal{D}_\pi(m),\\
J_5(q_0,m)&=&q_0J_3(q_0,m)-2J_4(q_0,m), \\
J_6(q_0,m)&=&\frac{1}{D+1}\left(\left(m^2-q_0^2\right)J_2(q_0,m)-\frac{m^2 q_0}{d}\mathcal{D}_\pi(m) \right),\\
J_7(q_0,m)&=&m^2 J_2(q_0,m)+(D+2)J_6(q_0,m).
\eea
We also define the derivative function
\bea
J^{(n)}_i(\qv,m)&=&\frac{\partial^{n}}{\partial( m^2)^n}J_i(\qv,m).
\eea

%%%
\section{Amplitudes for the diagrams}\label{Amp}

Throughout this work we use the normalization $\bar u(p)u(p)=2M_p$ and we define $\Delta=M_\Delta-M_N$. $\mt^2=\mpi^2-q^2 x(1-x)$ and the function $Z$ has been defined in \eq{Z}. We work in the rest frame where $v=(1,{\bf 0})$.

\subsection{Pion loops}

Here we collect the amplitudes of all the diagrams contributing to the proton polarizability through a loop of pions, represented in \fig{piL}, plus the ones with a crossed photon
lines or permutations, which are assumed to be implicit in the representation. For all the diagrams here we consider the overall factor $\mathcal A = 2 M_p \frac{g_{A}^2}{F_\pi^2} $. We assume a positive infinitesimal imaginary part for the propagators of $h_{14}-h_{19}$.
Diagrams with only 1 pion are zero due to the fact that we are working in the static limit. 

\bea
\mathcal M_1^{\mu\nu}&=&\mathcal{A}\, g^{\mu\nu}h_0(q^2,\qv),\\
\mathcal M_2^{\mu\nu}&=&\mathcal{A}\left\{h_1(q^2,\qv)(g^{\mu\nu}-v^\mu v^\nu)+h_2(q^2,\qv)i\epsilon^{ \mu\nu\alpha\beta}v_\alpha S_\beta\right\},\\
\mathcal M^{\mu\nu}_3&=&\mathcal{A}\left\{h_3(q^2,\qv) g^{\mu\nu}+h_4(q^2,\qv)q^\mu q^\nu +h_5(q^2,\qv) (q^\mu v^\nu+v^\mu q^\nu)+h_6(q^2,\qv) v^\mu v^\nu\right\},\\
\mathcal M_{4}^{\mu\nu}&=&\mathcal{A}\left\{h_7(q^2,\qv)g^{\mu\nu}+h_8(q^2,\qv)q^\mu q^\nu+h_9(q^2,\qv)v^\nu v^\mu+ h_{10}(q^2,\qv)(q^\mu v^\nu +q^\nu v^\mu)\right.\nonumber\\
&+&h_{13}(q^2,\qv) i (\epsilon^{\mu \lambda \alpha \beta} v^\nu-\epsilon^{\nu \lambda\alpha \beta} v^\mu)q_\lambda  S_\beta  v_\alpha +\left.h_{11}(q^2,\qv)i\epsilon^{\mu \nu \alpha \beta} S_\beta v_\alpha\right.\nn\\
&+&\left. h_{12}(q^2,\qv)i(\epsilon^{\mu \lambda \alpha\beta} q^\nu-\epsilon^{\nu\lambda \alpha \beta} q^\mu)q_\lambda S_\beta v_\alpha\right\},\\
\mathcal M_{5}^{\mu\nu}&=&\mathcal{A}\left\{h_{14}(q^2,\qv)v^\mu v^\nu+h_{15}(q^2,\qv)(q^\mu v^\nu +q^\nu v^\mu)+h_{16}(q^2,\qv) i (\epsilon^{\mu \lambda \alpha \beta} v^\nu-\epsilon^{\nu \lambda\alpha \beta} v^\mu)q_\lambda  S_\beta  v_\alpha\right\},\nn\\
\mathcal M_{6}^{\mu\nu}&=&\mathcal{A}h_{17}(q^2,\qv)v^\mu v^\nu ,\\
\mathcal M_{7}^{\mu\nu}&=&\mathcal{A}h_{18}(q^2,\qv)v^\mu v^\nu,\\
\mathcal M_{8}^{\mu\nu}&=&\mathcal{A}h_{19}(q^2,\qv)v^\mu v^\nu.
\eea
where the $h$ functions read:
\bea
h_0(q^2,\qv)&=&-J_0(0,m_\pi) -\mpi^2 J_0'(0,m_\pi),\\
h_1(q^2,\qv)&=&\frac{1}{2}\left(J_0(\qv,m_\pi^2)+J_0(-\qv,m_\pi^2)\right),\\
h_2(q^2,\qv)&=&-J_0(\qv,m_\pi^2)+J_0(-\qv,m_\pi^2),\\
h_3(q^2,\qv)&=&2 \int_0^1 dx(1-x) \left\{(D+1) \left(J_6''(\qv x,\tilde m^2)+J_6''(-\qv x,\tilde m^2)\right)\right.\nn\\
&-&\left.x^2 \vq^2 \left(J_2''(\qv x,\tilde m^2)+J_2''(-q_0 x,\tilde m^2)\right)\right\}\\
h_4(q^2,\qv)&=&\frac{1}{2} \int_0^1 dx (1-x) (2 x-1) \left\{(D (2 x-1)+6 x+1) \left(J_2''(\qv x,\tilde m^2)+J_2''(-q_0 x,\tilde m^2)\right)\right.\nn\\
&-&\left.(2 x-1) x^2 \vq^2 \left(J_0''(\qv x,\tilde m^2)+J_0''(-\qv x,\tilde m^2)\right)\right\},
\eea
\bea
h_5(q^2,\qv)&=&\intx(1-x)\left\{(-2 D x+D-2 x-1) \left(J_4''(\qv x,\tilde m^2)-J_4''(-\qv x,\tilde m^2)\right)\right.\nn\\
&+&\left.x (2 x-1) \left(x \vq^2 \left(J_1''(\qv x,\tilde m^2)-J_1''(-\qv x,\tilde m^2)\right)\right.\right.\nn\\
&-&\left.\left.2 q_0 \left(J_2''(\qv x,\tilde m^2)+J_2''(-\qv x,\tilde m^2)\right)\right)\right\},\\
h_6(q^2,\qv)&=&2 \intx(1-x)  \left\{(D-1) \left(J_7''(\qv x,\tilde m^2)+J_7''(-\qv x,\tilde m^2)\right)\right.\nn\\
&+&\left.x \left(-x\vq^2 \left(J_3''(\qv x,\tilde m^2)+J_3''(-\qv x,\tilde m^2)\right)+4 \qv \left(J_4''(\qv x,\tilde m^2)-J_4''(-\qv x,\tilde m^2)\right)\right)\right.\nn\\
&-&\left.2 \left(J_6''(\qv x,\tilde m^2)+J_6''(-\qv x,\tilde m^2)\right)\right\},\\
h_7(q^2,\qv)&=&-2\intx \left\{J_2'\left(\qv x,\tilde m^2\right)+J_2'\left(-\qv x,\tilde m^2\right)\right\},\\
h_8(q^2,\qv)&=&\intx x (1-2 x) \left\{J_0'\left(\qv x,\tilde m^2)\right)+J_0'\left(-\qv x,\tilde m^2\right)\right\},\\
h_{9}(q^2,\qv)&=&2\intx \left\{-\qv x \left(J_1'\left(\qv x,\tilde m^2\right)-J_1'\left(-\qv x,\tilde m^2\right)\right)+J_2'\left(\qv x,\tilde m^2\right)+J_2'\left(-\qv x,\tilde m^2\right)\right\},\nn\\
\\
h_{10}(q^2,\qv)&=&\intx x \left\{\frac{\qv}{2} (2 x-1)\left( J_0'\left(\qv x,\tilde m^2\right)+J_0'\left(-\qv x,\tilde m^2\right)\right)\right.\nn\\
&+&\left.J_1'\left(\qv x,\tilde m^2\right)- J_1'\left(-\qv x,\tilde m^2\right)\right\},\\
h_{11}(q^2,\qv)&=&4\intx \left\{J_2'\left(\qv x,\tilde m^2\right)-J_2'\left(-\qv x,\tilde m^2\right)\right\},\\
h_{12}(q^2,\qv)&=&- \intx x (1-2x) \left\{J_0'\left(\qv x,\tilde m^2\right)-J_0'\left(-\qv x,\tilde m^2\right)\right\}, \\
h_{13}(q^2,\qv)&=&-2 \intx x \left\{J_1'\left(\qv x,\tilde m^2\right)+J_1'\left(-\qv x,\tilde m^2\right)\right\},\\
h_{14}(q^2,\qv)&=&\frac{2}{\qv}\intx \left\{(D-1) (J_4'\left(\qv x,\mt^2\right)- J_4'\left(-\qv x,\mt^2\right))+\vq^2 (1-x) x\left( J_1'\left(\qv x,\mt^2\right)\right.\right.\nn\\
&-&\left.\left. J_1'\left(-\qv x,\mt^2\right)\right)-\qv (1-2 x) (J_2'\left(\qv x,\mt^2\right)+ J_2'\left(-\qv x,\mt^2\right))\right\},\\
h_{15}(q^2,\qv)&=&\frac{1}{2 \qv}\intx (1-2 x) \left\{(D+1) \left(J_2'\left(\qv x,\mt^2\right)+J_2'\left(-\qv x,\mt^2\right)\right)\right.\nn\\
&+&\left.\vq^2  x(1-x) \left(J_0'\left(\qv x,\mt^2\right)+ J_0'\left(-\qv x,\mt^2\right)\right)\right\},\\
h_{16}(q^2,\qv)&=&-\frac{2}{\qv}\intx \left\{J_2'(\qv x,\mt^2)-J_2'(-\qv x,\mt^2)\right\}\\
h_{17}(q^2,\qv)&=&-2\frac{D-1}{4}\frac{1}{\qv^2}\left(-2J_2\left(0,\mpi^2\right)+J_2\left(-\qv,\mpi^2\right)+J_2\left(\qv,\mpi^2\right)\right),\\
h_{18}(q^2,\qv)&=&3 \frac{D-1}{4}\frac{1}{\qv^2}\left(J_2\left(\qv,\mpi^2\right)+J_2\left(-\qv,\mpi^2\right)-\left(J_2\left(0,m^2\right)+J_2\left(0,\mpi^2\right)\right)\right),\\
h_{19}(q^2,\qv)&=&\frac{D-1}{4}\frac{1}{\qv}\left(\frac{1}{\qv}\left(J_2\left(\qv,\mpi^2\right)+J_2\left(-\qv,\mpi^2\right)-2J_2\left(0,\mpi^2\right)\right)\right)\nn\\.
\eea
and, for $D=4-\epsilon$ dimensions we obtain:
\bea
h_0(q^2,\qv)&=&\frac{3  m_\pi}{16 \pi }+\mathcal O(\epsilon),\\
h_1(q^2,\qv)&=&-\frac{ \sqrt{m_\pi^2-\qv^2}}{8 \pi }+\mathcal O(\epsilon),\\
h_2(q^2,\qv)&=&\frac{1}{4 \pi ^2}\qv\tL+\frac{1}{4 \pi ^2}\left(2 \sqrt{\mpi^2-\qv^2} \sin^{-1}\left(\frac{\qv}{\mpi}\right)-\qv\right)+\mathcal O(\epsilon),\\
h_3(q^2,\qv)&=&\frac{1 }{16 \pi }\left(\frac{ \left(6 \mpi^2 q^2-8\mpi^2 \qv^2-q^4\right)}{2 \vq^2\sqrt{\vq^2}}\mathcal{I}_{1}-\frac{\mpi }{\vq^2}\sqrt{1-\frac{\qv^2}{\mpi^2}} \left(2 \mpi^2-q^2+2 \qv^2\right)+\frac{\mpi \left(2 \mpi^2+q^2\right)}{\vq^2}\right)\nn\\
&+&\mathcal O(\epsilon),\\
h_4(q^2,\qv)&=&\frac{-1 }{16 \pi }\left(\frac{ \left(-6 \mpi^2 \left(q^2-2 \qv^2\right)+q^4+2 \qv^4\right) \left(4 \mpi^2 \left(\qv^2-q^2\right)+q^4\right)}{2 \vq^4\sqrt{\vq^2} \left(4 \mpi^2 {\bf q}^2+q^4\right)}\mathcal{I}_{1}\right.\nn\\
&+&\left.\frac{\mpi \left(16\mpi^4 \left(q^2-\qv^2\right)-2 m^2 \left(6 q^4-16 q^2 \qv^2+13 \qv^4\right)+q^2 \left(2 q^4-6 q^2 \qv^2+\qv^4\right)\right)}{\vq^4 \left(4 \mpi^2 \vq^2+q^4\right)}\right.\nn\\
&+&\left.\frac{ \left(\mpi \left(16 \mpi^4 \left(\qv^2-q^2\right)+\mpi^2 \left(10 \qv^4-4 q^2 \qv^2\right)+q^6+2 q^4\qv^2\right)\right)}{{\bf q}^4 \left(4 \mpi^2 {\bf q}^2+q^4\right)}\sqrt{1-\frac{\qv^2}{\mpi^2}}\right)+\mathcal O(\epsilon),\\
h_5(q^2,\qv)&=&\frac{1 }{16 \pi}\left(-\frac{\mpi \qv \left(16 \mpi^4\vq^2-6 \mpi^2 \qv^2 \left(q^2-2 \qv^2\right)+q^6+2 q^4 \qv^2\right)}{\vq^4 \left(4 \mpi^2\vq^2+q^4\right)}\right.\nn\\
&+&\left.\frac{\qv \left(\mpi^2 \left(10 \qv^2-4 q^2\right)+q^4+2 q^2 \qv^2\right)}{2 \vq^4\sqrt{\vq^2}}\mathcal{I}_{1}+\frac{ \left(\mpi\qv \left(16 \mpi^4\vq^2+\mpi^2 \left(14 q^2 \qv^2-8 q^4\right)+3 q^6\right)\right)}{\vq^4 \left(4 m^2 {\bf q}^2+q^4\right)}\right.\nn\\
&&\left.\sqrt{1-\frac{\qv^2}{\mpi^2}}\right)+\mathcal O(\epsilon),\\
h_6(q^2,\qv)&=&\frac{1}{16 \pi }\left(-\frac{ q^2 \left(-6 \mpi^2 \left(q^2-2\qv^2\right)+q^4+2 q^2 \qv^2\right)}{2  \vq^4\sqrt{\vq^2}}\mathcal{I}_{1}\right.\nn\\
&+&\left.\frac{\mpi \left(8 \mpi^4 \left(\qv^4-q^4\right)-2 \mpi^2 \left(q^6+2 q^4\qv^2-6 q^2 \qv^4\right)+q^8+2 q^6 \qv^2\right)}{ \vq^4 \left(4 \mpi^2  {\bf q}^2+q^4\right)}\right.\nn\\
&+&\left.\frac{ \left(\mpi \left(8 \mpi^4 \left(q^4-\qv^4\right)+\mpi^2 \left(-6 q^6+32 q^4 \qv^2-48 q^2 \qv^4+16 \qv^6\right)+q^8-8 q^6 \qv^2+4 q^4 \qv^4\right)\right)}{ {\bf q}^4 \left(4\mpi^2  \vq^2+q^4\right)}\right.\nn\\
&&\left.\sqrt{1-\frac{\qv^2}{\mpi^2}}\right)+\mathcal O(\epsilon),\\
h_7(q^2,\qv)&=&\frac{1}{16 \pi } \left(\frac{ \left(4 \mpi^2\vq^2+q^4\right)}{2 \vq^2\sqrt{\vq^2}}\mathcal{I}_{1}-\frac{\sqrt{1-\frac{\qv^2}{\mpi^2}} \left(\mpi \left(q^2-2 \qv^2\right)\right)}{\vq^2}-\frac{\mpi q^2}{{\bf q}^2}\right)+\mathcal O(\epsilon),
\eea
\bea
h_8(q^2,\qv)&=&\frac{1}{16 \pi } \left(\frac{ \left(4 \mpi^2 \vq^2+q^4+2 q^2 \qv^2\right)}{2 \vq^4\sqrt{\vq^2}}\mathcal{I}_{1}+\frac{3 \mpi q^2 \sqrt{1-\frac{\qv^2}{\mpi^2}}}{\vq^4}-\frac{\mpi \left(q^2+2 \qv^2\right)}{\vq^4}\right)+\mathcal O(\epsilon),\nn\\
\\
h_{9}(q^2,\qv)&=&\frac{1 }{16 \pi }\left(\frac{ \left(4 \mpi^2 q^2 \vq^2+q^6+2 q^4 \qv^2\right)}{2\vq^4\sqrt{\vq^2}}\mathcal{I}_{1}-\frac{\mpi \left(q^4-8 q^2 \qv^2+4 \qv^4\right)}{\vq^4}\sqrt{1-\frac{\qv^2}{m^2}}\right.\nn\\
&-&\left.\frac{\mpi q^2 \left(q^2+2 \qv^2\right)}{\vq^4}\right)+\mathcal O(\epsilon),\\
h_{10}(q^2,\qv)&=&\frac{1}{16 \pi } \left(-\frac{ \qv \left(4 \mpi^2\vq^2+q^2 \left(2 q^2+\qv^2\right)\right)}{2 \vq^4 \sqrt{\vq^2}}\mathcal{I}_{1}\right.\nn\\
&+&\left.\frac{\sqrt{1-\frac{\qv^2}{\mpi^2}} \left(\mpi \qv \left(\qv^2-4 q^2\right)\right)}{\vq^4}+\frac{\mpi\qv \left(2 q^2+\qv^2\right)}{\vq^4}\right)+\mathcal O(\epsilon),\\
h_{11}(q^2,\qv)&=&-\frac{1}{4 \pi ^2}\qv \tL+\frac{\qv}{4 \pi ^2}+\frac{-1}{2\pi ^2}\left(2\sqrt{\mt^2-\qv^2 x^2} \sin^{-1}\left(\frac{\qv x}{\sqrt{\mt^2}}\right)+\qv x \ln\left(\frac{\mt^2}{\mpi^2}\right)\right.\nn\\
&+&\mathcal O(\epsilon),\\
h_{12}(q^2,\qv)&=&\frac{1}{4\pi^2}\intx x (1-2x) \frac{ \sin ^{-1}\left(\frac{\qv x}{\sqrt{\mt^2}}\right)}{\sqrt{\mt^2-\qv^2 x^2}}+\mathcal O(\epsilon),\\
h_{13}(q^2,\qv)&=&\frac{-1}{8\pi^2}\tL-\frac{1}{8\pi^2}-\frac{1}{4\pi^2}\intx  x \left\{\ln\left(\frac{\mt^2}{\mpi^2}\right)-\frac{2 \qv x \sin ^{-1}\left(\frac{\qv x}{\sqrt{\mt^2}}\right)}{\sqrt{\mt^2-\qv^2 x^2}}\right\}+\mathcal O(\epsilon),\\
h_{14}(q^2,\qv)&=&\frac{1}{16 \pi}\frac{ q^2 }{  \left(\vq^2\right)^{3/2}}\left(2 \mpi^2-q^2\right)\mathcal{I}_{1}-\frac{1}{8 \pi}\frac{\mpi}{  \vq^2}\left(2 \mpi^2- q^2\right)\nn\\
&+&\frac{1}{8 \pi}\frac{\mpi }{  \vq^2}\left(2 \mpi^2+q^2-2\qv^2\right) \sqrt{1-\frac{\qv^2}{\mpi^2}}+\mathcal O(\epsilon),\\
h_{15}(q^2,\qv)&=&-\frac{1}{32 \pi}\frac{  2 \mpi^2-q^2}{  \vq^2\sqrt{\vq^2}}\qv \mathcal{I}_{1}-\frac{1}{16 \pi}\frac{\mpi}{\qv \vq^2}\left(2 \mpi^2-\qv^2\right)\left(\sqrt{1-\frac{\qv^2}{\mpi^2}}-1\right)+\mathcal O(\epsilon),\\
h_{16}(q^2,\qv)&=&\frac{1}{8 \pi ^2}\tL-\frac{1}{8 \pi ^2}+\frac{1}{4 \pi ^2 }\intx\left\{ x \ln \left(\frac{\mt^2}{\mpi^2}\right)+\frac{2}{\qv} \sqrt{\mt^2-\qv^2 x^2} \sin ^{-1}\left(\frac{\qv x}{\sqrt{\mt^2}}\right)\right\}\nn\\
&+&\mathcal O(\epsilon),\\
h_{17}(q^2,\qv)&=&-\frac{1}{8 \pi  }\frac{\mpi^3}{\qv^2}\left(1-\left(1-\frac{\qv^2}{\mpi^2}\right)^{3/2}\right)+\mathcal O(\epsilon),\\
h_{18}(q^2,\qv)&=&\frac{3  }{16 \pi  }\frac{\mpi^3}{\qv^2}\left(1-\left(1-\frac{\qv^2}{\mpi^2}\right)^{3/2}\right)+\mathcal O(\epsilon),\\
h_{19}(q^2,\qv)&=&\frac{1}{16 \pi  }\frac{\mpi^3}{\qv^2}\left(1-\left(1-\frac{\qv^2}{\mpi^2}\right)^{3/2}\right)+\mathcal O(\epsilon).
\eea
These expressions agree with Eqs. (81)-(84) of \cite{Hemmert:1996rw} when $\qv=0$ and $\epsilon\cdot v =0$.

We have explicitly checked that our result is gauge invariant through the following relations between the $h$'s:
\bea
&&h_2(q^2,\qv)+h_{11}(q^2,\qv)+q^2 h_{12}(q^2,\qv)+\qv (h_{13}(q^2,\qv)+h_{16}(q^2,\qv))=0\,,\\
&&h_0(q^2,\qv)+h_1(q^2,\qv)+h_3(q^2,\qv)+h_{7}(q^2,\qv)+\qv\left(h_{10}(q^2,\qv)+h_{15}(q^2,\qv)\right)\nn\\
&&\hspace{8.25cm}+q^2\left(h_{4}(q^2,\qv)+h_{8}(q^2,\qv)\right)=0\,,\\
&&-\frac{\qv^2}{q^2}\left(-h_{1}(q^2,\qv)+h_{6}(q^2,\qv)+h_{9}(q^2,\qv)+h_{14}(q^2,\qv)+h_{17}(q^2,\qv)+h_{18}(q^2,\qv)+h_{19}(q^2,\qv)\right)\nn\\
&&+h_0(q^2,\qv)+h_1(q^2,\qv)+h_3(q^2,\qv)+h_{7}(q^2,\qv)+q^2\left(h_{4}(q^2,\qv)+h_{8}(q^2,\qv)\right)=0
\,.
\eea

\subsection{Pion loops which include a $\Delta$ excitation}

Here we collect the amplitudes of all the diagrams contributing to the proton polarizability with a $\Delta$ particle and through a loop of pions, represented in \fig{D1L}, plus the ones with a crossed photon lines or permutations, which are assumed to be implicit in the representation. For all the diagrams here we consider the overall factor $\mathcal{A}=-\frac{8}{3}M_p\frac{g_{\pi N \Delta}^2}{F_\pi^2}$. We take a positive infinitesimal imaginary part for the propagators of $h_{14}^\Delta-h_{19}^\Delta$.

\bea
\mathcal M_{\Delta\pi 1}^{\mu\nu}&=&\mathcal{A}\,g^{\mu\nu}h_0^\Delta(q^2,\qv),\\
\mathcal M_{\Delta\pi 2}^{\mu\nu}&=&\mathcal{A}\left\{(g^{\mu\nu}-v^\mu v^\nu)h_1^\Delta (q^2,\qv)+i\epsilon^{\mu\nu\alpha\beta}v_\alpha S_\beta h_2^\Delta (q^2,\qv)\right\},\\
\mathcal M_{\Delta\pi3}^{\mu\nu}&=&\mathcal{A}\left\{g^{\mu\nu}h_3^\Delta (q^2,\qv)+q^\mu q^\nu h_4^{\Delta}(q^2,\qv)+(q^\mu v^\nu+v^\mu q^\nu)h_5^\Delta(q^2,\qv)+v^\mu v^\nu h_6^\Delta (q^2,\qv)\right\},\nn\\
\\
\mathcal M_{\Delta\pi4}^{\mu\nu}&=&\mathcal{A}\left\{g^{\mu\nu}h_7^\Delta (q^2,\qv)+q^\mu q^\nu h_8^{\Delta}(q^2,\qv)+(q^\mu v^\nu+v^\mu q^\nu)h_{10}^\Delta(q^2,\qv)+v^\mu v^\nu h_{9}^\Delta (q^2,\qv)\right.\nn\\
&+&\left.i\epsilon^{\mu\nu\alpha\beta}v_\alpha S_\beta h_{11}^\Delta(q^2,\qv)+iv_\alpha S_\beta q_\lambda(\epsilon^{\mu\lambda\alpha\beta}q^\nu-\epsilon^
{\nu\lambda\alpha\beta}q^\mu)h_{12}^\Delta(q^2,\qv)\right.\nn\\
&+&\left.iv_\alpha S_\beta q_\lambda(\epsilon^{\mu\lambda\alpha\beta}v^\nu-\epsilon^{\nu\lambda\alpha\beta}v^\mu)
h_{13}^\Delta(q^2,\qv)\right\},\\
\mathcal M_{\Delta\pi5}^{\mu\nu}&=&\mathcal{A}\left\{v^\mu v^\nu h_{14}^\Delta (q^2,\qv)+(q^\mu v^\nu+v^\mu q^\nu)h_{15}^\Delta(q^2,\qv)\right.\nn\\
&+&\left.iv_\alpha S_\beta q_\lambda(\epsilon^{\mu\lambda\alpha\beta}v^\nu-\epsilon^{\nu\lambda\alpha\beta}v^\mu)
h_{16}^\Delta(q^2,\qv)\right\},\\
\mathcal M_{\Delta\pi 6}^{\mu\nu}&=&\mathcal{A}\left\{v^\mu v^\nu h_{17}^\Delta (q^2,\qv)\right\},\\
\mathcal M_{\Delta\pi 7}^{\mu\nu}&=&\mathcal{A}\left\{v^\mu v^\nu h_{18}^\Delta (q^2,\qv)\right\},\\
\mathcal M_{\Delta\pi 8}^{\mu\nu}&=&\mathcal{A}\left\{v^\mu v^\nu h_{19}^\Delta (q^2,\qv)\right\}.
\eea
where in terms of the master integrals:
\bea
h_0^\Delta(q^2,\qv)&=&-2(D-2)J_2'(-\Delta ,\mpi^2),\\
h_1^{\Delta}(q^2,\qv)&=&\frac{D-2}{ D-1}\left(J_0\left(\qv-\Delta ,\mpi^2\right)+J_0\left(-\qv-\Delta ,\mpi^2\right)\right),\\
h_2^\Delta (q^2,\qv)&=&\frac{-2 }{D-1}\left(J_0\left(\qv-\Delta ,m_\pi^2\right)-J_0\left(-\qv-\Delta ,m_\pi^2\right)\right),\\
h_3^\Delta (q^2,\qv)&=&4\frac{D-2}{D-1} \intx(1-x) \left\{-\vq^2x^2 \left(J_2''\left(\qv x-\Delta ,\mt^2\right)+J_2''\left(-\qv x-\Delta ,\mt^2\right)\right)\right.\nn\\
&+&\left.(D+1) \left(J_6''\left(\qv x-\Delta ,\mt^2\right)+J_6''\left(-\qv x-\Delta ,\mt^2\right)\right)\right\},\\
h_4^\Delta (q^2,\qv)&=& \frac{D-2}{D-1}\intx (1-x) (2 x-1) \left\{-\vq^2 x^2 (2 x-1)(J_0''\left(\qv x-\Delta ,\mt^2\right)+J_0''\left(-\qv x-\Delta ,\mt^2\right))\right.\nn\\
&+&\left. (4 x(D+1)-(1+2x)(D-1)) \left(J_2''\left(\qv x-\Delta ,\mt^2\right)+J_2''\left(-\qv x-\Delta ,\mt^2\right)\right)\right\},\\
h_5^\Delta (q^2,\qv)&=&2\frac{D-2}{D-1}\intx\left\{(-\vq^2x ^2(1-2 x)  \left(J_1''\left(\qv x-\Delta ,\mt^2\right)-J_1''\left(-\qv x-\Delta ,\mt^2\right)\right)\right.\nn\\
&+&\left.2 \qv x (1-2 x)\left(J_2''\left(\qv x-\Delta ,\mt\right)+J_2''\left(-\qv x-\Delta ,\mt^2\right)\right)\right)\nn\\
&+&\left.(D-1-2 (D+1) x) \left(J_4''\left(\qv x-\Delta ,\mt^2\right)-J_4''\left(-\qv x-\Delta ,\mt^2\right)\right)\right\},\\
h_6^\Delta (q^2,\qv)&=&4 \frac{D-2}{D-1}\intx(1-x) \left\{x \left(-\vq^2 x \left(J_3''\left(\qv x-\Delta ,\mt^2\right)+ J_3''\left(-\qv x-\Delta ,\mt^2\right)\right)\right.\right.\nn\\
&+&\left.\left.4\qv \left(J_4''\left(\qv x-\Delta ,\mt^2\right)-J_4''\left(-\qv x-\Delta ,\mt^2\right)\right)\right)\right.\nn\\
&-&\left.2 \left(J_6''\left(\qv x-\Delta ,\mt^2\right)+J_6''\left(-\qv x-\Delta ,\mt^2\right)\right)\right\},\\
h_7^\Delta (q^2,\qv)&=&-4\frac{D-2 }{D-1}\intx\left\{J_2'\left( \qv x-\Delta ,\mt^2\right)+J_2'\left(-\qv x-\Delta ,\mt^2\right)\right\},\\
h_8^\Delta (q^2,\qv)&=&2\frac{D-2 }{D-1}\intx(1-2 x) x \left\{J_0'\left( \qv x-\Delta ,\mt^2\right)+J_0'\left(-\qv x-\Delta ,\mt^2\right)\right\},\\
h_{9}^\Delta(q^2,\qv)&=&4\frac{D-2}{D-1}\intx \left\{J_2'\left(\qv x-\Delta,\mt^2\right)+J_2'\left(-\qv x-\Delta ,\mt^2\right)\right.\nn\\
&-&\left.\qv x (J_1'(\qv x-\Delta,\mt^2)-J_1'(-\qv x-\Delta,\mt^2))\right\},\\
h_{10}^\Delta(q^2,\qv)&=&\frac{D-2}{D-1}\intx x \left\{2\left( J_1'\left(\qv x-\Delta ,\mt^2\right)- J_1'\left(-\qv x-\Delta ,\mt^2\right)\right)\right.\nn\\
&-&\left.(1-2 x) \qv\left(J_0'\left(\qv x-\Delta ,\mt^2\right)+J_0'\left(-\qv x-\Delta,\mt^2\right)\right) \right\},\\
h_{11}^\Delta(q^2,\qv)&=&\frac{8 }{D-1}\intx \left\{J_2'\left(\qv x-\Delta,\mt^2\right)-J_2'\left(-\qv x-\Delta ,\mt^2\right)\right\},
\eea
\bea
h_{12}^\Delta(q^2,\qv)&=&-\frac{2 }{D-1}\intx x (1-2 x) \left\{J_0'\left(\qv x-\Delta ,\mt^2\right)-J_0'\left(-\qv x-\Delta,\mt^2\right)\right\},\\
h_{13}^\Delta(q^2,\qv)&=& -\frac{4 }{D-1}\intx x\left\{J_1'\left(\qv x-\Delta ,\mt^2\right)+J_1'\left(-\qv x-\Delta ,\mt^2\right)\right\},\\
 h_{14}^\Delta (q^2,\qv)&=&4 \frac{D-2}{D-1 } \frac{1}{\qv}\intx \left\{(D-1)\left( J_4'\left(\qv x-\Delta,\mt^2\right)-J_4'\left(-\qv x-\Delta,\mt^2\right)\right)\right.\nn\\
 &-&\left.(1-2 x) \qv\left( J_2'\left(\qv x-\Delta ,\mt^2\right)+J_2'\left(-\qv x-\Delta ,\mt^2\right) \right)\right.\nn\\
 &+&\left.(1-x) x \vq^2\left( J_1'\left(\qv x-\Delta ,\mt^2\right)- J_1'\left(-\qv x-\Delta ,\mt^2\right) \right)\right\},\\
 h_{15}^\Delta (q^2,\qv)&=& \frac{D-2 }{D-1 }\frac{1}{ \qv}\intx(1-2 x) \left\{(D+1)\left( J_2'\left(\qv x-\Delta,\mt^2\right)+ J_2'\left(-\qv x-\Delta ,\mt^2\right)\right)\right.\nn\\
 &+&\left.\vq^2(1-x) x \left(J_0'\left(\qv x-\Delta,\mt^2\right) +J_0'\left(-\qv x-\Delta ,\mt^2\right) \right)\right\},\\
 h_{16}^\Delta (q^2,\qv)&=&-\frac{2}{D-1}\frac{2}{\qv}\intx \left\{J_2'(\qv x-\Delta,\mt^2)-J_2'(-\qv x-\Delta,\mt^2)\right\}\\
h_{17}(q^2,\qv)&=&-2\frac{D-1}{4}\frac{1}{\qv^2}\left(-2J_2\left(0,\mpi^2\right)+J_2\left(-\qv,\mpi^2\right)+J_2\left(\qv,\mpi^2\right)\right),\\
 h _{17}^\Delta (q^2,\qv)&=&2 \frac{D-2}{D-1}\left(\frac{1-D}{2}\frac{1}{\qv^2}\left(-2 J_2\left(-\Delta ,\mpi^2\right)+J_2\left(-\qv-\Delta ,\mpi^2\right)+J_2\left(\qv-\Delta , \mpi^2\right)\right)\right),\nn\\
 \\
h _{18}^\Delta (q^2,\qv)&=&2 \frac{D-2}{D-1} \frac{D-1}{4}\frac{3}{\qv^2}\left(J_2\left(\qv-\Delta ,\mpi^2\right)+J_2\left(-\qv-\Delta ,\mpi^2\right)-2J_2\left(-\Delta ,\mpi^2\right)\right),\\
h_{19}^\Delta (q^2,\qv)&=&2 \frac{D-2}{D-1} \frac{D-1}{4}\frac{1}{\qv^2}\left(J_2\left(\qv-\Delta ,\mpi^2\right)+J_2\left(-\qv-\Delta ,\mpi^2\right)-2J_2\left(-\Delta ,\mpi^2\right)\right).
\eea
These results, in the limit $q^2=0$ and in the gauge where $\epsilon\cdot v=0$, agree with Eqs. (89)-(92) of \cite{Hemmert:1996rw}.

Now, expanding in $D=4-\epsilon$ we get
\bea
h_0^\Delta (q^2,\qv)&=&-\frac{1 }{4 \pi ^2}\Delta \, \tL-\frac{1}{2 \pi ^2}\mpi \mathcal{Z}\left(\frac{\Delta }{\mpi}\right),\\
h_1^\Delta (q^2,\qv)&=&\frac{1}{6 \pi ^2}\Delta  \tL+\frac{1}{18 \pi ^2}\left(3 \mpi \left(\mathcal{Z}\left(\frac{\Delta-\qv}{\mpi}\right)+\mathcal{Z}\left(\frac{\Delta +\qv}{\mpi}\right)\right)-2 \Delta \right)+\mathcal O (\epsilon),\\
h_2^\Delta (q^2,\qv)&=&\frac{1}{6 \pi ^2}\qv \tL+\frac{1}{6 \pi ^2}\left(\mpi \left(\mathcal{Z}\left(\frac{\Delta +\qv}{\mpi}\right)-\mathcal{Z}\left(\frac{\Delta-\qv}{\mpi}\right)\right)-\frac{5 \qv}{3}\right)+\mathcal O (\epsilon),\\
h_3^\Delta (q^2,\qv)&=&\frac{5 \Delta  \tL}{12 \pi ^2}-\frac{\Delta }{9 \pi ^2}+\frac{ 1}{6 \pi ^2}\intx (1-x)\left\{5 \Delta  \ln \left(\frac{\mt^2}{\mpi^2}\right)+\sqrt{\mt^2} \left(\left(5-\frac{\vq^2 x^2}{\mt^2-(\Delta +\qv x)^2}\right)\right.\right.\nn\\
&&\left.\left. \mathcal{Z}\left(\frac{\Delta +\qv x}{\sqrt{\mt^2}}\right)+\left(5-\frac{\vq^2 x^2}{\mt^2-(\Delta -\qv x)^2}\right) \mathcal{Z}\left(\frac{\Delta -\qv x}{\sqrt{\mt^2}}\right)\right)\right\}+\mathcal O(\epsilon),
\eea
\bea
h_4^\Delta (q^2,\qv)&=&\frac{1}{24 \pi ^2}\intx(1-x) (2 x-1) \left\{x^2 (2 x-1)\frac{\vq^2 }{\mt^2} \left(\frac{\Delta +\qv x}{\mt^2-(\Delta +\qv x)^2}+\frac{\Delta -\qv x}{\mt^2-(\Delta -\qv x)^2}\right)\right.\nn\\
&-&\left.\frac{\sqrt{\mt^2}  }{\mt^2-(\Delta +\qv x)^2}\left(3 \left(1-\frac{14 x}{3}\right)-\frac{\vq^2 x^2 (2 x-1)}{\mt^2-(\Delta +\qv x)^2}\right)\mathcal{Z}\left(\frac{\Delta +\qv x}{\sqrt{\mt^2}}\right)\right.\nn\\
&-&\left.\frac{\sqrt{\mt^2} }{\mt^2-(\Delta -\qv x)^2}\left(3 \left(1-\frac{14 x}{3}\right)-\frac{\vq^2 x^2 (2 x-1)}{\mt^2-(\Delta -\qv x)^2} \right)\mathcal{Z}\left(\frac{\Delta -\qv x}{\sqrt{\mt^2}}\right)\right\}+\mathcal{O}(\epsilon),\\
h_5^\Delta (q^2,\qv)&=&\frac{1}{12 \pi ^2}\intx\left\{(2 x-1) x^2 \left(\frac{\vq^2}{\mt^2-(\Delta -\qv x)^2}-\frac{\vq^2}{\mt^2-(\Delta +\qv x)^2}\right)\right.\nn\\
&-&\left.\frac{\sqrt{\mt^2} }{\mt^2-(\Delta +\qv x)^2}\left(\frac{x^2 (2 x-1) \vq^2 (\Delta +\qv x)}{\mt^2-(\Delta +\qv x)^2}-\qv x (5 (1-2 x)-4 x)\right.\right.\nn\\
&+&\left.\Delta  (4 x+3 (2 x-1))\frac{}{}\right) \mathcal{Z}\left(\frac{\Delta +\qv x}{\sqrt{\mt^2}}\right)+\frac{\sqrt{\mt^2} }{\mt^2-(\Delta -\qv x)^2}\left(\frac{x^2 (2 x-1)\vq^2 (\Delta -\qv x)}{\mt^2-(\Delta -\qv x)^2}\right.\nn\\
&+&\left.\left.\qv x (5 (1-2 x)-4 x)+\Delta  (4 x+3 (2 x-1))\frac{}{}\right) \mathcal{Z}\left(\frac{\Delta -\qv x}{\sqrt{\mt^2}}\right)\right\}+\mathcal{O}(\epsilon),\\
h_6^\Delta (q^2,\qv)&=&-\frac{1}{6 \pi ^2}\Delta  \tL+\frac{\Delta }{9 \pi ^2}+\frac{1}{6 \pi ^2}\intx(1-x) \left\{-2 \Delta  \ln \left(\frac{\mt^2}{\mpi^2}\right)\right.\nn\\
&+&\left.\vq^2 x^2 \left(\frac{\Delta +\qv x}{\mt^2-(\Delta +\qv x)^2}+\frac{\Delta -\qv x}{\mt^2-(\Delta -\qv x)^2}\right)+\sqrt{\mt^2} \left(\frac{\vq^2 x^2 (\Delta +\qv x)^2}{\left(\mt^2-(\Delta +\qv x)^2\right)^2}\right.\right.\nn\\
&+&\left.\left.\frac{4 \qv x (\Delta +\qv x)+\vq^2 x^2}{\mt^2-(\Delta +\qv x)^2}-2\right) \mathcal{Z}\left(\frac{\Delta +\qv x}{\sqrt{\mt^2}}\right)+\sqrt{\mt^2}\left(\frac{\vq^2 x^2 (\Delta -\qv x)^2}{\left(\mt^2-(\Delta -\qv x)^2\right)^2}\right.\right.\nn\\
&+&\left.\left.\frac{\vq^2 x^2-4 \qv x (\Delta -\qv x)}{\mt^2-(\Delta -\qv x)^2}-2\right) \mathcal{Z}\left(\frac{\Delta -\qv x}{\sqrt{\mt^2}}\right)\right\}+\mathcal{O}(\epsilon),\\
h_7^\Delta (q^2,\qv)&=&-\frac{1}{3 \pi ^2}\Delta  \tL+\frac{2 \Delta }{9 \pi ^2}\nn\\
&-&\frac{1}{3 \pi ^2}\intx \left\{\Delta  \ln \left(\frac{\mt^2}{\mpi^2}\right)+\sqrt{\mt^2} \left(\mathcal{Z}\left(\frac{\Delta -\qv x}{\sqrt{\mt^2}}\right)+\mathcal{Z}\left(\frac{\Delta +\qv x}{\sqrt{\mt^2}}\right)\right)\right\}+\mathcal{O}(\epsilon),\\
h_8^\Delta (q^2,\qv)&=&\frac{1}{6 \pi ^2}\intx  (1-2 x) x\, \sqrt{\mt^2} \left\{\frac{\mathcal{Z}\left(\frac{\Delta +\qv x}{\sqrt{\mt^2}}\right)}{\mt^2-(\Delta +\qv x)^2}+\frac{\mathcal{Z}\left(\frac{\Delta -\qv x}{\sqrt{\mt^2}}\right)}{\mt^2-(\Delta -\qv x)^2}\right\}+\mathcal{O}(\epsilon),\\
h_9^\Delta (q^2,\qv)&=&\frac{1}{3 \pi ^2}\Delta  \tL-\frac{2 \Delta }{9 \pi ^2}+\frac{1}{3 \pi ^2}\intx \left\{\Delta  \ln \left(\frac{\mt^2}{\mpi^2}\right)+\sqrt{\mt^2} \left(\left(1-\frac{x (\qv (\Delta +\qv x))}{\mt^2-(\Delta +\qv x)^2}\right)\right. \right.\nn\\
&&\left.\left.\mathcal{Z}\left(\frac{\Delta +\qv x}{\sqrt{\mt^2}}\right)+\left(\frac{x (\qv (\Delta -\qv x))}{\mt^2-(\Delta -\qv x)^2}+1\right) \mathcal{Z}\left(\frac{\Delta -\qv x}{\sqrt{\mt^2}}\right)\right)\right\}+\mathcal{O}(\epsilon),\\
h_{10}^\Delta (q^2,\qv)&=&\frac{1}{12 \pi ^2}\intx \sqrt{\mt^2} x \left\{\frac{(2 \Delta +\qv (4 x-1)) \mathcal{Z}\left(\frac{\Delta +\qv x}{\sqrt{\mt^2}}\right)}{\mt^2-(\Delta +\qv x)^2}+\frac{(\qv(4 x-1)-2 \Delta ) \mathcal{Z}\left(\frac{\Delta -\qv x}{\sqrt{\mt^2}}\right)}{\mt^2-(\Delta -\qv x)^2}\right\}\nn\\
&+&\mathcal{O}(\epsilon),
\eea
\bea
h_{11}^\Delta (q^2,\qv)&=&-\frac{\qv \tL}{6 \pi ^2}+\frac{5 \qv}{18 \pi ^2}-\frac{1}{3 \pi ^2}\intx \left\{\qv x \ln \left(\frac{\mt^2}{\mpi^2}\right)-\sqrt{\mt^2} \left(\mathcal{Z}\left(\frac{\Delta -\qv x}{\sqrt{\mt^2}}\right)\right.\right.\nn\\
&-&\left.\left.\mathcal{Z}\left(\frac{\Delta +\qv x}{\sqrt{\mt^2}}\right)\right)\right\}+\mathcal{O}(\epsilon),\\
h_{12}^\Delta (q^2,\qv)&=&-\frac{1}{12 \pi ^2}\intx  (1-2x) x\, \sqrt{\mt^2} \left\{\frac{\mathcal{Z}\left(\frac{\Delta -\qv x}{\sqrt{\mt^2}}\right)}{\mt^2-(\Delta -\qv x)^2}-\frac{\mathcal{Z}\left(\frac{\Delta +\qv x}{\sqrt{\mt^2}}\right)}{\mt^2-(\Delta +\qv x)^2}\right\}+\mathcal{O}(\epsilon),\\
h_{13}^\Delta (q^2,\qv)&=&-\frac{\tL}{12 \pi ^2}-\frac{1}{36 \pi ^2}-\frac{1}{6 \pi ^2}\intx x \left\{\ln \left(\frac{\mt^2}{\mpi^2}\right)\right.\nn\\
&-&\left.\sqrt{\mt^2} \left(\frac{(\Delta +\qv x) \mathcal{Z}\left(\frac{\Delta +\qv x}{\sqrt{\mt^2}}\right)}{\mt^2-(\Delta +\qv x)^2}+\frac{(\Delta -\qv x) \mathcal{Z}\left(\frac{\Delta -\qv x}{\sqrt{\mt^2}}\right)}{\mt^2-(\Delta -\qv x)^2}\right)\right\}+\mathcal{O}(\epsilon),\\
h_{14}^\Delta (q^2,\qv)&=&\frac{\Delta\tL }{\pi^2}+\frac{1}{3\pi ^2}\intx\left\{(-1+8 x) \Delta  \ln\left(\frac{\mt^2}{\mpi^2}\right)\right.\nn\\
&+&\left.\frac{\sqrt{\mt^2}}{\qv}\left(\left(3\Delta -\qv(1-5x)-\frac{\vq^2(1-x) x (\qv x+\Delta ) }{(\Delta +\qv x)^2-\mt^2}\right)\mathcal{Z}\left(\frac{-\qv x-\Delta }{\sqrt{\mt^2}}\right)\right.\right.\nn\\
&&\left.\left.-\left(3\Delta +\qv(1-5x)-\frac{\vq^2 (1-x) x (-\qv x+\Delta ) }{(\Delta -\qv x)^2-\mt^2}\right)\mathcal{Z}\left(\frac{\qv x-\Delta }{\sqrt{\mt^2}}\right)\right)\right\}+\mathcal{O}(\epsilon),\\
h_{15}^\Delta (q^2,\qv)&=&\frac{1}{24 \pi ^2\qv}\intx (1-2 x)\left\{ 10 \Delta  \ln \left(\frac{\mt^2}{\mpi^2}\right)+2 \sqrt{\mt^2} \left(\left(\frac{\vq^2 (1-x) x}{\mt^2-(\Delta +\qv x)^2}+5\right)\right.\right.\nn\\
 &&\left.\left. \mathcal{Z}\left(\frac{\Delta +\qv x}{\sqrt{\mt^2}}\right)+\left(\frac{\vq^2 (1-x) x}{\mt^2-(\Delta -\qv x)^2}+5\right) \mathcal{Z}\left(\frac{\Delta -\qv x}{\sqrt{\mt^2}}\right)\right)\right\}+\mathcal{O}(\epsilon),\\
h_{16}^\Delta (q^2,\qv)&=&\frac{\tL}{12\pi^2}-\frac{5}{36\pi^2}+\frac{1}{6\pi^2}\intx \left\{x\ln\left(\frac{\mt^2}{\mpi^2}\right)+\frac{\sqrt{\mt^2}}{\qv}\left(\mathcal{Z}\left(\frac{\Delta+\qv x}{\sqrt{\mt^2}}\right)-\mathcal{Z}\left(\frac{\Delta-\qv x}{\sqrt{\mt^2}}\right)\right)\right\}\nn\\
&+&\mathcal{O}(\epsilon),\\
h_{17}^\Delta (q^2,\qv)&=&\frac{-1}{6\pi ^2}\left(- 3 \tL \Delta +2\Delta -\frac{\mpi }{ \qv^2}\left(\left((\qv+\Delta )^2-\mpi^2\right) \mathcal{Z}\left(\frac{\Delta +\qv}{\mpi}\right)\right.\right.\nn\\
&+&\left.\left.\left((\qv-\Delta )^2-\mpi^2\right) \mathcal{Z}\left(\frac{-\Delta +\qv}{\mpi}\right)-2\left(-\mpi^2+\Delta ^2\right) \mathcal{Z}\left(-\frac{\Delta }{\mpi}\right)\right)\right)+\mathcal{O}(\epsilon),\\
 h _{18}^\Delta (q^2,\qv)&=&\frac{1}{4\pi ^2}\left(- 3 \tL \Delta +2\Delta -\frac{\mpi }{ \qv^2}\left(\left((\qv+\Delta )^2-\mpi^2\right) \mathcal{Z}\left(\frac{\Delta +\qv}{\mpi}\right)\right.\right.\nn\\
&+&\left.\left.\left((\qv-\Delta )^2-\mpi^2\right) \mathcal{Z}\left(\frac{-\Delta +\qv}{\mpi}\right)-2\left(-\mpi^2+\Delta ^2\right) \mathcal{Z}\left(-\frac{\Delta }{\mpi}\right)\right)\right)+\mathcal{O}(\epsilon),\\
h_{19}^\Delta (q^2,\qv)&=&\frac{1}{12\pi ^2}\left(- 3 \tL \Delta +2\Delta -\frac{\mpi }{ \qv^2}\left(\left((\qv+\Delta )^2-\mpi^2\right) \mathcal{Z}\left(\frac{\Delta +\qv}{\mpi}\right)\right.\right.\nn\\
&+&\left.\left.\left((\qv-\Delta )^2-\mpi^2\right) \mathcal{Z}\left(\frac{-\Delta +\qv}{\mpi}\right)-2\left(-\mpi^2+\Delta ^2\right) \mathcal{Z}\left(\frac{\Delta }{\mpi}\right)\right)\right)+\mathcal{O}(\epsilon).
\eea

We have explicitly checked that our result is gauge invariant through the following relations between the $h^\Delta$'s:
\bea
&&h_2^\Delta(q^2,\qv)+h_{11}^\Delta(q^2,\qv)+q^2 h_{12}^\Delta(q^2,\qv)+\qv (h_{13}^\Delta(q^2,\qv)+h_{16}^\Delta(q^2,\qv))=0,\\
&&h_0^\Delta(q^2,\qv)+h_1^\Delta(q^2,\qv)+h_3^\Delta(q^2,\qv)+h_{7}^\Delta(q^2,\qv)+\qv\left(h_{10}^\Delta(q^2,\qv)+h_{15}^\Delta(q^2,\qv)\right)\nn\\
&&\hspace{8.75cm}+q^2\left(h_{4}^\Delta(q^2,\qv)+h_{8}^\Delta(q^2,\qv)\right)=0,\\
&&-\frac{\qv^2}{q^2}\left(-h_{1}^\Delta(q^2,\qv)+h_{6}^\Delta(q^2,\qv)+h_{9}^\Delta(q^2,\qv)+h_{14}^\Delta(q^2,\qv)+h_{17}^\Delta(q^2,\qv)+h_{18}^\Delta(q^2,\qv)+h_{19}^\Delta(q^2,\qv)\right)\nn\\
&&+h_0^\Delta(q^2,\qv)+h_1^\Delta(q^2,\qv)+h_3^\Delta(q^2,\qv)+h_{7}^\Delta(q^2,\qv)+q^2\left(h_{4}^\Delta(q^2,\qv)+h_{8}^\Delta(q^2,\qv)\right)=0,
\eea
which are equivalent to:

\bea
&&h_2(q^2,\qv-\Delta)+h_{11}(q^2,\qv-\Delta)+q^2 h_{12}(q^2,\qv-\Delta)+\qv (h_{13}(q^2,\qv)+h_{16}(q^2,\qv-\Delta))=0,\nn\\
\\
&&h_0(q^2,\qv-\Delta)+h_1(q^2,\qv-\Delta)+h_3(q^2,\qv-\Delta)+h_{7}(q^2,\qv-\Delta)+\qv\left(h_{10}(q^2,\qv-\Delta)\right.\nn\\
&&\hspace{2.25cm}\left.+h_{15}(q^2,\qv-\Delta)\right)+q^2\left(h_{4}(q^2,\qv-\Delta)+h_{8}(q^2,\qv-\Delta)\right)=0,\\
&&-\frac{\qv^2}{q^2}\left(-h_{1}(q^2,\qv-\Delta)+h_{6}(q^2,\qv-\Delta)+h_{9}(q^2,\qv-\Delta)+h_{14}(q^2,\qv-\Delta)+h_{17}(q^2,\qv-\Delta)\right.\nn\\
&&\left.+h_{18}(q^2,\qv-\Delta)+h_{19}(q^2,\qv-\Delta)\right)+h_0(q^2,\qv-\Delta)+h_1(q^2,\qv-\Delta)\nn\\
&&+h_3(q^2,\qv-\Delta)+h_{7}(q^2,\qv-\Delta)+q^2\left(h_{4}(q^2,\qv-\Delta)+h_{8}(q^2,\qv-\Delta)\right)=0.
\eea

\end{document}